\documentclass[a4paper,10pt]{article}

\usepackage{amsfonts}
\usepackage{latexsym}
\usepackage{epsfig}

\usepackage{amssymb}

\usepackage{graphicx}
\usepackage{oldgerm}
\usepackage{theorem}

\def\vlambda{\mib{\lambda}}
\def\vkappa{\mib{\kappa}}
\def\vomega{\mib{\omega}}

\def\R{\mathbb{R}}
\def\C{\mathbb{C}}
\def\N{\mathbb{N}}

\def\W{\mathbb{W}}
\def\B{\mathbb{B}}

\def\x{{\bf x}}
\def\y{{\bf y}}
\def\z{{\bf z}}
\def\u{{\bf u}}
\def\a{{\bf a}}
\def\0{{\bf 0}}

\def\X{{\bf X}}
\def\Y{{\bf Y}}

\usepackage{theorem}

\renewcommand{\thesection}{\Roman{section}}

\newcommand{\mib}[1]{\mbox{\boldmath $#1$}}

\theorembodyfont{\itshape}
\newtheorem{thm}{Theorem}
\newtheorem{lem}[thm]{Lemma}
\newtheorem{cor}[thm]{Corollary}
\newtheorem{prop}[thm]{Proposition}

\newcommand{\qed}{\hbox{\rule[-2pt]{3pt}{6pt}}}
\begin{document}
\pagestyle{plain}
\vskip 0.3cm

\noindent
{\bf \Large{Symmetry of matrix-valued stochastic processes\\
and noncolliding diffusion particle systems}}

\vskip 0.3cm

\noindent
Makoto Katori
\footnote{Electronic mail: katori@phys.chuo-u.ac.jp}\\
{\it Department of Physics,
Faculty of Science and Engineering,\\
Chuo University, 
Kasuga, Bunkyo-ku, 
Tokyo 112-8551, Japan}

\vskip 0.3cm

\noindent
Hideki Tanemura
\footnote{Electronic mail: tanemura@math.s.chiba-u.ac.jp}\\
{\it Department of Mathematics and Informatics,
Faculty of Science,\\
Chiba University, 
1-33 Yayoi-cho, Inage-ku,
Chiba 263-8522, Japan}

\vskip 0.5cm

\noindent  
As an extension of the theory of Dyson's Brownian motion models 
for the standard Gaussian random-matrix ensembles,
we report a systematic study of hermitian matrix-valued
processes and their eigenvalue processes associated with
the chiral and nonstandard random-matrix ensembles.
In addition to the noncolliding Brownian motions, we introduce
a one-parameter family of temporally homogeneous noncolliding
systems of the Bessel processes and a two-parameter family of
temporally inhomogeneous noncolliding systems of Yor's
generalized meanders and show that all of the ten classes
of eigenvalue statistics in the Altland-Zirnbauer classification
are realized as particle distributions 
in the special cases of these diffusion particle systems. 
As a corollary of each equivalence
in distribution of a temporally inhomogeneous eigenvalue
process and a noncolliding diffusion process,
a stochastic-calculus proof of a version of
the Harish-Chandra (Itzykson-Zuber) formula
of integral over unitary group is established.



\normalsize

\section{\large INTRODUCTION}

It is interesting to consider today mathematical-physical
sequences of the two classic papers
\cite{Dys62b} and \cite{Dys62a} by Dyson of random matrix
theory, which appeared sequentially in the same volume
of the journal in 1962. In one of them  \cite{Dys62b}, 
following the early work of Wigner, 
he gave a logical foundation for
his classification scheme of random-matrix ensembles
based on the group representation theory of Weyl
and established the standard (Wigner-Dyson) random matrix
theory for the three ensembles
called the Gaussian unitary, orthogonal, and symplectic
ensembles (GUE, GOE, and GSE). 
He introduced in the other paper \cite{Dys62a} the
hermitian matrix-valued Brownian motions,
which are associated with these Gaussian random-matrix
ensembles, and studied the stochastic processes of
eigenvalues of the matrix-valued processes.
Combining the standard perturbation theory of the quantum
mechanics and a simple but essential consideration 
of the scaling of Brownian motions, he generally proved
that the obtained eigenvalue processes are identified with
the one-dimensional systems of Brownian particles
with the repulsive two-body forces proportional 
to the inverse of distances between particles. 
These processes are now called
Dyson's Brownian motion models 
$\Y(t)=(Y_{1}(t), Y_{2}(t), \cdots, Y_{N}(t))$
described by the stochastic differential equations
\begin{equation}
d Y_{i}(t)= dB_{i}(t)+ \frac{\beta}{2}
\sum_{1 \leq j \leq N, j \not=i}
\frac{1}{Y_{i}(t)-Y_{j}(t)} dt,
\quad t \in [0, \infty), 1 \leq i \leq N,
\label{eqn:Dyson1}
\end{equation}
with $\beta=1,2,4$ for GOE, GUE and GSE, respectively,
where $B_{i}(t), 1 \leq i \leq N$ are independent
one-dimensional standard Brownian motions.
Dyson's classification scheme has been extended. 
In addition to the standard three random-matrix ensembles,
their {\it chiral versions} (chGUE, chGOE, and chGSE) 
were studied in the particle physics of QCD 
associated with consideration of the gauge groups
and quantum numbers called flavors 
\cite{VZ93,Ver94,JSV96,SV98}.
After that extension, Altland and Zirnbauer introduced
more four ensembles called the classes
C, CI, D, and DIII for the solid-state
physics of mesoscopic systems considering the
particle-hole symmetry, which plays an important
role in the Bogoliubov-de Gennes framework of the
BCS mean-field theory of superconductivity\cite{AZ96,AZ97}.
These totally ten Gaussian ensembles are systematically
argued by Zirnbauer \cite{Zir96} based on Cartan's classification
scheme of symmetric spaces \cite{Hel78} and Efetov's supersymmetry
theory \cite{Efe97}.

One consequence of a combination
of the two papers by Dyson may be to give a
systematic study of matrix-valued diffusion processes
({\it i.e.} diffusion processes in groups or algebraic
spaces) and perform the classification of 
eigenvalue processes as generalization of Dyson's
Brownian motion models.
This line has been taken by Bru \cite{Bru89,Bru91},
Grabiner \cite{Gra99},
K\"onig and O'Connell \cite{KO01} and others,
and one of the purpose of the present paper
is to clarify the relationship between 
statistics of (nonstandard) random matrix theory 
and stochastic processes of interacting diffusion particles
in the type of Dyson's Brownian motion models
studied in the probability theory.
We will claim in Sec.II that the matrix-valued processes
called the Wishart process by Bru \cite{Bru91} 
and the Laguerre process by K\"onig and O'Connell \cite{KO01} 
are the stochastic versions of chGOE and chGUE, 
respectively, in the sense of Dyson \cite{Dys62a},
and derive in Sec.III the diffusion processes 
describing the eigenvalue statistics of the classes C
and D of Altland and Zirnbauer, following 
Bru's matrix-version of the stochastic calculus
based on the Ito rule for differentials.

Due to the strong repulsive forces in the processes of
the types of Dyson's Brownian motion models,
particle collisions are suppressed. Impossibility
of collision may be generally proved by the same argument
as Bru, who showed that the collision time between
two eigenvalues of the Wishart process is 
infinite ($\tau = + \infty$ a.s.) \cite{Bru89}.
For the $\beta=2$ (GUE) case of Dyson's Brownian motion
model (\ref{eqn:Dyson1}), 
if $\Y(0) \in \W_{N}^{\rm A}$ then $\Y(t) \in \W_{N}^{\rm A}$
for all $t >0$ with probability 1, where
$\W_{N}^{\rm A}$ denotes the Weyl chamber of type 
$\mbox{A}_{N-1}$;
$\W_{N}^{\rm A}=\{ \x \in \R^N ; 
x_{1} < x_{2} < \cdots < x_{N}\}$.
Using the Karlin-McGregor formula \cite{KM59a,KM59b} 
the transition density of the absorbing
Brownian motion in $\W_{N}^{\rm A}$ from the state
$\x$ at time $s$ to the state $\y$
at time $t (> s)$ is given by the determinant
\begin{equation}
f^{\rm A}(t-s, \y|\x)=\det_{1 \leq i, j \leq N}
\Bigg[ G^{\rm A}(t-s, y_{j}|x_{i}) \Bigg], \quad
\x, \y \in \W_{N}^{\rm A}, 
\label{eqn:fNA}
\end{equation}
where each element is the Gaussian heat-kernel
$G^{\rm A}(t, y|x)=e^{-(x-y)^2/2t}/\sqrt{2 \pi t}.$
Grabiner \cite{Gra99} pointed out that 
the transition probability density of the process
(\ref{eqn:Dyson1}) with $\beta=2$ is given by
$$
 p^{\rm A}(s, \x; t, \y)=
\frac{1}{h^{\rm A}(\x)} 
f^{\rm A}(t-s, \y|\x) h^{\rm A}(\y),
$$
where 
$h^{\rm A}(\x)=\prod_{1 \leq i < j \leq N} (x_{j}-x_{i})$.
Since $h^{\rm A}(\x)$ is a strictly positive harmonic function
in $\W_{N}^{\rm A}$, this is regarded as the $h$-transform 
in the sense of Doob \cite{Doo84}, and it implies
that the eigenvalue process of GUE is realized as the
noncolliding Brownian motions
({\it i.e.} the $h$-transform of an absorbing Brownian motion in 
the Weyl chamber of type $\mbox{A}_{N-1}$).
K\"onig and O'Connell also showed that the eigenvalue
process of the Laguerre process, which corresponds
to chGUE, is realized as the noncolliding system
of the squared Bessel processes \cite{KO01}.
In Sec.IV, we show that the
eigenvalue processes of random matrices
in the symmetry classes C and D of Altland and Zirnbauer
are realized as the noncolliding system
of the {\it Brownian motions with an absorbing
wall at the origin} \cite{KTNK03} 
({\it i.e.} the $h$-transform of an absorbing Brownian motion in 
the Weyl chamber of type $\mbox{C}_{N}$) and as
the noncolliding system of the
{\it reflecting Brownian motions}
({\it i.e.} the $h$-transform of an absorbing Brownian motion 
in the Weyl chamber of type $\mbox{D}_{N}$), respectively.
These three kinds of systems are discussed
as special cases of a family of noncolliding systems
of diffusion particles with one parameter $\nu > -1$,
in which each particle is following the $d=2(\nu+1)$-dimensional 
Bessel process defined by the transition probability density
\cite{BS96,RY98}
\begin{eqnarray}
G^{(\nu)}(t, y|x) &=& \frac{y^{\nu+1}}{x^{\nu}}
\frac{1}{t} e^{-(x^2+y^2)/2t} 
I_{\nu} \left( \frac{xy}{t} \right) \quad
\mbox{for} \quad x>0, y \geq 0, \nonumber\\
G^{(\nu)}(t, y|0) &=& \frac{y^{2\nu+1}}
{2^{\nu} \Gamma(\nu+1) t^{\nu+1}} 
e^{-y^2/2t} \quad
\mbox{for} \quad y \geq 0,
\label{eqn:Bessel1}
\end{eqnarray}
where $\Gamma$ denotes the Gamma function and
$I_{\nu}$ is the modified Bessel function;
$I_{\nu}(z)=\sum_{n=0}^{\infty} 
(z/2)^{2n+\nu}/\{\Gamma(n+1) \Gamma(\nu+n+1) \}$.

How can we realize other six eigenvalue
processes in Altland-Zirnbauer's ten classes
of random-matrix ensembles as well by noncolliding systems
of diffusion processes ?
In our previous papers \cite{KT02,KT03a} we considered the
situation that the noncolliding condition is imposed 
not forever but for a finite time-interval $(0, T]$ to define the
temporally inhomogeneous noncolliding Brownian motions
$\X(t)=(X_{1}(t), X_{2}(t), \cdots, X_{N}(t))$.
Of course, we can see that $\X(t) \to \Y(t)$ in distribution
as $T \to \infty$. We observed for the finite time-interval
$t \in [0,T]$ that, if we set $\X(0)=\Y(0)=\0$
with $\0=(0, 0, \cdots, 0) \in \R^{N}$, then
\begin{equation}
P( \X(\cdot) \in d{\bf w})
= \frac{ C[{\rm A}] T^{\psi[{\rm A}]}}
{C[{\rm A}'] h^{\rm A} ({\bf w}(T))}
P( \Y(\cdot) \in d {\bf w}),
\label{eqn:Imhof}
\end{equation}
where $C[{\rm A}]=(2 \pi)^{N/2} \prod_{i=1}^{N} \Gamma(i)$,
$C[{\rm A}']=2^{N/2} \prod_{i=1}^{N} \Gamma(i/2)$,
and $\psi[{\rm A}]=N(N-1)/4$.
This is regarded as a multivariate version of the Imhof relation 
in the probability theory \cite{Imh84}, since it implies 
the absolute continuity in distribution of
the temporally homogeneous process $\Y(t)$ and the
inhomogeneous process $\X(t)$ in $[0,T]$, but 
from the viewpoint of random matrix theory the 
important consequence of this equality is the fact that
the process $\X(t)$ exhibits a transition
in distribution from the eigenvalue statistics of
GUE to that of GOE and thus the GOE distribution
is realized at the final time $t=T$.
In Sec.V, we develop this argument by
replacing the Brownian motions $X_{i}(t), 1 \leq i \leq N$
by the {\it generalized meanders} with two parameters
$(\nu, \kappa), \nu > -1, \kappa \in [0, 2(\nu+1))$, 
introduced as the temporally inhomogeneous diffusions associated 
with the Bessel process by Yor \cite{Yor92}, whose 
transition probability density is given by
\begin{equation}
G^{(\nu,\kappa)}_{T}(s,x; t,y)
= \frac{1}{h^{(\nu,\kappa)}_{T}(s,x)}
G^{(\nu)}(t-s, y|x) h^{(\nu,\kappa)}_{T}(t,y)
\label{eqn:meander}
\end{equation}
for $0 \leq s < t \leq T, x, y \geq 0 $ with
$h^{(\nu,\kappa)}_{T}(t ,x)
= \int_0^\infty dz \ G^{(\nu)}(T-t, z|x) z^{-\kappa}$.
By choosing the two parameters $(\nu, \kappa)$ appropriately,
this family of noncolliding systems of 
generalized meanders provides such diffusion processes
that exhibit the transitions from chGUE
to chGOE and from the class C to the class CI.
We will also consider the processes, in which the noncolliding 
condition collapses at the final time $t=T$ in the ways that
all particles collide simultaneously or
only pairwise collisions occur.
In the special cases in the latter situation,
we have the processes showing the transitions
from GUE to GSE, from chGUE to chGSE, and
from the class D to the class DIII. 

The present study of the temporally inhomogeneous noncolliding 
diffusion processes gives two kinds of byproducts.
(i) Topology of path-configurations of our processes 
on the spatio-temporal plane $\R \times [0,T]$ is 
determined by the conditions at $t=0$ and $t=T$. 
We will be able to discuss the
topology of random directed polymer networks \cite{deG68,EG95}
using the random matrix theory.
Such correspondence between the topology of path-configurations
and random-matrix ensembles is recently used by
Sasamoto and Imamura to analyze one-dimensional
polynuclear growth models \cite{SI03}.
(ii) A variety of versions of Harish-Chandra (Itzykson-Zuber)
formulae of integrals over unitary groups \cite{HC57,IZ80} 
are derived as corollaries of the equivalence in distribution
of the eigenvalue processes of matrix-valued processes
and noncolliding diffusion processes.
Other remarks are given in Sec.VI.

\section{\large BRU'S THEOREM}
\subsection{Hermitian matrix-valued stochastic processes}

We denote the space of $N \times N$ hermitian matrices
by ${\cal H}(N)$, the group of $N \times N$
unitary matrices by ${\rm U}(N)$, and 
the group of $N \times N$
real orthogonal matrices by ${\rm O}(N)$.
We also use the notations 
${\cal S}(N)$ and ${\cal A}(N)$ for
the spaces of $N \times N$
real symmetric and real antisymmetric matrices, respectively.
We consider complex-valued processes 
$\xi_{ij}(t) \in \C, 1 \leq i, j \leq N, t \in [0, \infty),$
with the condition $\xi_{ji}(t)^{*}=\xi_{ij}(t)$,
and define the matrix-valued processes 
by $\Xi(t)=(\xi_{ij}(t))_{1 \leq i, j \leq N} \in {\cal H}(N)$.
We denote by 
$U(t)=(u_{ij}(t))_{1 \leq i, j \leq N}$ 
the family of unitary matrices which diagonalize $\Xi(t)$ so that
$$
U(t)^{\dagger} \Xi (t) U(t)=\Lambda(t)
={\rm diag}\{\lambda_{1}(t), \lambda_{2}(t), \cdots,
\lambda_{N}(t) \},
$$
where $\{\lambda_{i}(t)\}_{i=1}^{N}$ 
are eigenvalues of $\Xi(t)$ and we assume their increasing order
\begin{equation}
\lambda_{1}(t) \leq \lambda_{2}(t) \leq \cdots 
\leq \lambda_{N}(t).
\label{eqn:condition}
\end{equation}
Define $\Gamma_{ij}(t), 1 \leq i, j \leq N$, by
$\Gamma_{ij}(t) dt = 
(U(t)^{\dagger} d \Xi(t) U(t))_{ij}
(U(t)^{\dagger} d \Xi(t) U(t))_{ji}$,
where $d \Xi(t)=(d \xi_{ij})_{1 \leq i, j \leq N}$.
We denote by ${\bf 1}(\omega)$ the indicator function:
${\bf 1}(\omega)=1$ if the condition $\omega$ is satisfied, and
${\bf 1}(\omega)=0$ otherwise.
The following theorem is proved for the stochastic
process of eigenvalues $\vlambda(t)=(\lambda_{1}(t), \lambda_{2}(t),
\cdots, \lambda_{N}(t))$.

\begin{thm}
\label{thm:BruThm}
Assume that $\xi_{ij}(t), 1 \leq i < j \leq N$ are
continuous semimartingales.
The process $\vlambda(t)=(\lambda_{1}(t), \lambda_{2}(t),
\cdots, \lambda_{N}(t))$ satisfies 
the stochastic differential equations
$$
d \lambda_{i}(t)=dM_{i}(t)+d J_{i}(t), \quad
1 \leq i \leq N,
$$
where $M_{i}(t)$ is the martingale with quadratic variation
$
  \langle M_{i} \rangle_{t}=
  \int_{0}^{t} \Gamma_{ii}(s) ds
$
and $J_{i}(t)$ is the process with finite variation given by
$$
dJ_{i}(t) = \sum_{j=1}^{N} \frac{1}{\lambda_{i}(t)-\lambda_{j}(t)}
{\bf 1}(\lambda_{i}(t) \not= \lambda_{j}(t)) \Gamma_{ij}(t) dt
+ d \Upsilon_{i}(t)
$$
where $d \Upsilon_{i}(t)$ is the finite-variation part of
$(U(t)^{\dagger} d \Xi(t) U(t))_{ii}$.
\end{thm}
\vskip 0.5cm

Since this theorem is obtained by simple generalization
of Theorem 1 in Bru \cite{Bru89}, we call it 
Bru's theorem here.
A key point to derive the theorem is applying
the Ito rule for differentiating the product of
matrix-valued semimartingales:
If $X$ and $Y$ are $N \times N$ matrices
with semimartingale elements,
then
$$
d( X^{\dagger} Y)= (dX)^{\dagger} Y + X^{\dagger} (d Y) +
(dX)^{\dagger} (dY).
$$

\subsection{Four Basic Examples}

Let $\N=\{0,1,2, \cdots \}$ and assume $\nu \in \N$.
Let $B_{ij}(t)$, $\widetilde{B}_{ij}(t)$, 
$1\le i \leq N+\nu, 1 \leq j \leq N$
be independent one-dimensional standard Brownian motions. 
For $1 \leq i, j \leq N$ we set
$$
s_{ij}(t)
=
\left\{
   \begin{array}{ll}
      \displaystyle{
      \frac{1}{\sqrt{2}} B_{ij}(t)}, & 
   \mbox{if} \ i < j, \\
        & \\
        B_{ii}(t), & \mbox{if} \ i=j, \\
        & \\
    \displaystyle{
     \frac{1}{\sqrt{2}} B_{ji}(t)}, &
    \mbox{if} \ i > j, \\
   \end{array}\right. 
   \quad {\rm and } \quad
a_{ij}(t)
=
\left\{
   \begin{array}{ll}
      \displaystyle{
      \frac{1}{\sqrt{2}} \widetilde{B}_{ij}(t),
      } & 
   \mbox{if} \ i < j, \\
        & \\
    0, & \mbox{if} \ i=j,\\
        & \\
     \displaystyle{
      -\frac{1}{\sqrt{2}} \widetilde{B}_{ji}(t),
      } & 
   \mbox{if} \ i > j. \\
   \end{array}\right.
$$
Here we show four basic examples of hermitian matrix-valued processes
and applications of Theorem \ref{thm:BruThm}.

\begin{description}
\item{(i)} \quad
The first example of hermitian matrix-valued process is defined by
$$
\Xi(t) = (\xi_{ij}(t))_{1 \leq i, j \leq N}
=(s_{ij}(t)+\sqrt{-1}a_{ij}(t))_{1 \leq i, j \leq N},
\quad t \in [0, \infty).
$$
By definition 
$d\xi_{ij}(t) d\xi_{k \ell}(t)=\delta_{i \ell} \delta_{j k} dt,
\,1 \leq i, j, k, \ell \leq N,$
and thus $\Gamma_{ij}(t)=1$. Therefore
$\vlambda(t)$ solves the equations of Dyson's Brownian motion
model (\ref{eqn:Dyson1}) with $\beta=2$.

\item{(ii)} \quad
The second example is given by
$$
\Xi(t)=( s_{ij}(t))_{1 \leq i, j \leq N} \in {\cal S}(N),
\quad t \in [0, \infty).
$$
In this case
$
  d\xi_{ij}(t) d\xi_{k \ell}(t) =
(\delta_{i \ell} \delta_{j k}+\delta_{i k}\delta_{j \ell} ) dt/2,
\, 1 \leq i, j, k, \ell \leq N,
$
and thus
$
\Gamma_{ij}(t)dt=(1+\delta_{ij}) dt/2,
\, 1 \leq i, j \leq N.
$
Then $\vlambda(t)$ solves (\ref{eqn:Dyson1}) with $\beta=1$.

\item{(iii)} \quad
We consider an $(N+\nu) \times N$ matrix-valued process by
$M(t)=(B_{ij}(t)+\sqrt{-1} 
\widetilde{B}_{ij})_{1 \leq i \leq N+\nu, 1 \leq j \leq N}$
and define the $N \times N$ hermitian matrix-valued process by
\begin{equation}
   \Xi(t)= M(t)^{\dagger} M(t), \quad t \in [0, \infty).
\label{eqn:Laguerre1}
\end{equation}
Since the matrix $\Xi(t)$ is positive definite,
the eigenvalues are nonnegative.
By definition we see that
the finite-variation part of $d \xi_{ij}(t)$ is 
$2 (N+\nu) \delta_{ij} dt$ and
$d \xi_{ij}(t) d \xi_{k \ell}(t) =
2 (\xi_{i \ell}(t) \delta_{j k}
+\xi_{k j}(t) \delta_{i \ell}) dt,
\, 1 \leq i, j, k, \ell \leq N$,
which imply that $d \Upsilon_{i}(t)=2(N+\nu) dt$ and
$\Gamma_{ij}(t) = 2 (\lambda_{i}(t)+\lambda_{j}(t)),
\, 1 \leq i, j\leq N$.
Since $\langle M_{i} \rangle_{t}=
\int_{0}^{t} 4 \lambda_{i}(s) ds$,
the stochastic differential equations for $\vlambda(t)$ are given by
\begin{equation}
d\lambda_{i}(t)= 2 \sqrt{\lambda_{i}(t)} dB_{i}(t)
+ \beta \left\{ (N+\nu)+
\sum_{1 \leq j \leq N: j \not=i} \frac{\lambda_{i}(t)+\lambda_{j}(t)}
{\lambda_{i}(t)-\lambda_{j}(t)} \right\} dt,
\quad 1 \leq i \leq N,
\label{eqn:LW0}
\end{equation}
with $\beta=2$.

\item{(iv)} \quad
Set $B(t)=(B_{ij}(t))_{1 \leq i \leq N+\nu, 1 \leq j \leq N}$
and define 
\begin{equation}
   \Xi(t)= B(t)^{T} B(t) \in {\cal S}(N),
   \quad t \in [0, \infty).
\label{eqn:Wishart1}
\end{equation}
We see that the finite-variation part of $d \xi_{ij}(t)$
is $(N+\nu) \delta_{ij} dt$ and
$d \xi_{ij}(t) d \xi_{k \ell}(t) =
(\xi_{ik}(t) \delta_{j \ell}+\xi_{i \ell}(t) \delta_{jk}
+\xi_{j k}(t) \delta_{i \ell}+\xi_{j \ell}(t) \delta_{i k}) dt,
\, 1 \leq i, j, k, \ell \leq N$.
Then $d \Upsilon_{i}(t)=(N+\nu) dt$ and
$\Gamma_{ij}(t) = (\lambda_{i}(t)+\lambda_{j}(t)) (1+\delta_{ij}),
\, 1 \leq i, j\leq N$.
The equations for $\vlambda(t)$ are given by (\ref{eqn:LW0})
with $\beta=1$.
\end{description}

The process (\ref{eqn:Wishart1}) was called the Wishart process
and studied as matrix generalization of 
squared Bessel process by Bru \cite{Bru91}.
K\"onig and O'Connell \cite{KO01} called the process 
(\ref{eqn:Laguerre1}) the Laguerre process 
and studied its eigenvalue process (\ref{eqn:LW0}) with $\beta=2$.

\subsection{Relation with the standard and chiral 
random matrix theories}

Here we assume that $B_{ij}(0)=\widetilde{B}_{ij}(0)=0$
for all $1 \leq i \leq N+\nu, 1 \leq j \leq N$, and thus
the initial distribution of $\Xi(t)$ is the pointmass
on an $N \times N$ zero matrix $O$;
$\mu(\Xi \in \cdot ; 0)=\delta_{O}$.
In this case the distributions of $\Xi(t)$'s are related
with those studies in 
the standard (Wigner-Dyson) random matrix theory \cite{Meh91}
and the chiral random matrix theory \cite{VZ93,Ver94,JSV96,SV98}.

\vskip 0.5cm
\noindent{(i)} {\it Example (i) and GUE.} \quad
For GUE with variance $\sigma^2=t$ of random 
matrices in the space
${\cal H}(N) \cong \R^{d[{\rm A}]}$ with $d[{\rm A}]=N^2$, 
the probability density of eigenvalues $\vlambda$ 
in the condition (\ref{eqn:condition}) is given as \cite{Meh91}
$$
q^{\rm GUE}(\vlambda; t)
=\frac{t^{-d[{\rm A}]/2}}{C[{\rm A}]}
\exp \left\{ -\frac{|\vlambda|^2}{2t} \right\} 
h^{\rm A}(\vlambda)^2,
$$
where $|\vlambda|^2=\sum_{i=1}^{N} \lambda_{i}^2$.
For (\ref{eqn:Dyson1}) with $\beta=2$,
$p^{\rm A}(0, \0; t, \vlambda)=q^{\rm GUE}(\vlambda; t),
\, t > 0$.

\vskip 0.5cm
\noindent{(ii)} {\it Example (ii) and GOE.} \quad
The probability density of eigenvalues $\vlambda$
with the condition (\ref{eqn:condition}) is given as 
\cite{Meh91}
$$
q^{\rm GOE}(\vlambda; t)
= \frac{t^{-d[{\rm A}']/2}}{C[{\rm A}']} 
\exp \left\{ - \frac{|\vlambda|^2}{2t} \right\} 
h^{\rm A}(\vlambda)
$$
for GOE with variance $\sigma^2=t$ in
${\cal S}(N) \cong \R^{d[{\rm A}']}$,
$d[{\rm A}']=N(N+1)/2$.
If we denote by
$p^{\rm A'}(s, \vlambda; t, \vlambda')$
the transition probability density of the 
process (\ref{eqn:Dyson1}) with $\beta=1$
from $\vlambda$ at time $s$ to
$\vlambda'$ at time $t (> s)$, then
$p^{\rm A'}(0, \0; t, \vlambda)
=q^{\rm GOE}(\vlambda; t), \, t >0$.

\vskip 0.5cm
\noindent{(iii)} {\it Example (iii) and chiral GUE.} \quad
We denote by ${\cal M}(N+\nu, N; \C)$ 
and ${\cal M}(N+\nu, N; \R)$ the spaces of 
$(N+\nu) \times N$ complex and real matrices, respectively.
We see that 
${\cal M}(N+\nu, N; \C) \cong \R^{2N(N+\nu)}$ and write
its volume element as ${\cal V}(d M), 
M \in {\cal M}(N+\nu, N; \C)$.
The chiral Gaussian unitary ensemble (chGUE) 
with variance $t$ is the ensemble of matrices
$M \in {\cal M}(N+\nu, N; \C)$ with the 
probability density
\begin{equation}
\mu_{\nu}^{\rm chGUE}(M; t)
= \frac{t^{-N(N+\nu)/2}}{(2 \pi)^{N(N+\nu)}} \exp \left\{
-\frac{1}{2t} {\rm Tr} M^{\dagger} M \right\}
\label{eqn:Laguerre4}
\end{equation}
with respect to ${\cal V}(d M)$.
It is known \cite{Hua63} 
that any matrix $M \in {\cal M}(N+\nu, N; \C)$ has family of pairs
$(U, V), U \in {\rm U}(N+\nu), V \in {\rm U}(N)$,
which transform $M$ as $M = U^{\dagger} K V$,
where  $K \in {\cal M}(N+\nu,N; \R)$ is in the form 
$$
  K = \left( \matrix{ \widehat{K} \cr O} \right)
\quad
\mbox{with} \quad
\widehat{K}={\rm diag}
\{\kappa_{1}, \kappa_{2}, \cdots, \kappa_{N}\}, \quad
\kappa_{i} \geq 0, 1 \leq i \leq N,
$$
and the $\nu \times N$ zero matrix $O$.
We assume that $U$ and $V$ are chosen so that
\begin{equation}
0 \leq \kappa_{1} \leq \kappa_{2} \leq \cdots \leq \kappa_{N}.
\label{eqn:condition2}
\end{equation}
The matrices $(U, K, V)$ can be regarded as ``polar coordinates"
in the space ${\cal M}(N+\nu, N; \C)$. 
We have
$M^{\dagger} M = V^{\dagger} \Lambda V,$
where $\Lambda={\rm diag}\{\lambda_{1}, \lambda_{2}, \cdots,
\lambda_{N}\}$ with the relations
$\lambda_{i}=\kappa_{i}^2, \, 1 \leq i \leq N$.
Then $\vkappa=(\kappa_{1}, \kappa_{2}, \cdots, \kappa_{N})$
is a set of nonnegative square roots of the eigenvalues
of $M^{\dagger} M$.
Let $d \mu(U,V)$ be the Haar measure of
the space ${\rm U}(N+\nu) \times {\rm U}(N)$ normalized as
$\int_{{\rm U}(N+\nu) \times {\rm U}(N)} d\mu(U,V)=1$
and $d \vkappa=\prod_{i=1}^{N} d \kappa_{i}$.
Then we can show that
\begin{equation}
{\cal V}(dM)=\frac{(2 \pi)^{N(N+\nu)}}{C_{\nu}}
h^{((2\nu+1)/2)}(\vkappa)^2 d \vkappa d\mu(U,V),
\label{eqn:chGUEJ}
\end{equation}
where
$C_{\nu}=2^{N(N+\nu-1)} \prod_{i=1}^{N}
\{ \Gamma(i) \Gamma(i+\nu) \}$
and
$$
h^{(\alpha)}(\vkappa)=\prod_{1 \leq i<j \leq N}
(\kappa_{j}^2-\kappa_{i}^2) \prod_{k=1}^{N} \kappa_{k}^{\alpha}.
$$
For any pair of unitary matrices
$U \in {\rm U}(N+\nu)$ and $V \in {\rm U}(N)$,
the probability $\mu_{\nu}^{\rm chGUE}(M;t) {\cal V}(dM)$ 
is invariant under the automorphism
$M \to U^{\dagger}MV$.
By integrating over $d\mu(U,V)$, we obtain
the probability density of $\vkappa$
with the condition (\ref{eqn:condition2}) as
\cite{VZ93,Ver94,JSV96,SV98}
$$
q_{\nu}^{\rm chGUE}(\vkappa; t) 
= \frac{t^{-N(N+\nu)}}{C_{\nu}}
\exp \left\{ - \frac{|\vkappa|^2}{2t} \right\} 
h^{((2\nu+1)/2)}(\vkappa)^2.
$$

K\"onig and O'Connell \cite{KO01} studied the process
(\ref{eqn:LW0}) with $\beta=2$
as a multivariate version of squared Bessel process.
Here we consider the multivariate version of Bessel process
by extracting the square roots of eigenvalues
$\lambda_{i}(t) \geq 0$ of
$\Xi(t)=M(t)^{\dagger} M(t)$.
Setting $\kappa_{i}(t)=\sqrt{\lambda_{i}(t)} \geq 0,
1 \leq i \leq N$ in (\ref{eqn:LW0}) with $\beta=2$
and applying the Ito rule for differentials, 
we find that $\vkappa(t)$ solves 
the stochastic differential equations 
\begin{equation}
d Z_{i}(t)=d B_{i}(t)+ \frac{\beta}{2}
\left[
\frac{\gamma}{Z_{i}(t)}
+ \sum_{j: j \not=i} 
\left\{ \frac{1}{Z_{i}(t)-Z_{j}(t)}
+\frac{1}{Z_{i}(t)+Z_{j}(t)} \right\} \right] dt,
\quad 1 \leq i \leq N,
\label{eqn:basic1}
\end{equation}
with $(\beta, \gamma)=(2,(2 \nu+1)/2)$.
If we denote the transition probability density of this process by
$p^{(\nu)}(s, \, \cdot \, ; t, \, \cdot \,)$ 
for $0 \leq s < t < \infty$, then
\begin{equation}
p^{(\nu)}(0, \0; t, \vkappa)
=q_{\nu}^{\rm chGUE}(\vkappa; t), \quad t >0.
\label{eqn:Laguerre6}
\end{equation}

\vskip 0.5cm
\noindent{(iv)} {\it Example (iv) and chiral GOE.} \quad
We can see
${\cal M}(N+\nu, N; \R) \cong \R^{N(N+\nu)}$.
The chiral Gaussian orthogonal ensemble (chGOE) 
with variance $t$ is the ensemble of matrices
$B \in {\cal M}(N+\nu, N; \R) \subset
{\cal M}(N+\nu, N; \C)$ with the 
probability density
\begin{equation}
\mu_{\nu}^{\rm chGOE}(B; t)
=\frac{t^{-N(N+\nu)/2}}{(2 \pi)^{N(N+\nu)/2}} \exp \left\{
-\frac{1}{2t} {\rm Tr} B^{T} B \right\}
\label{eqn:Wishart4}
\end{equation}
with respect to the volume element 
${\cal V}'(dB)$ of ${\cal M}(N+\nu, N; \R)$.
We can show that
\begin{equation}
{\cal V}'(dB)=\frac{(2\pi)^{N(N+\nu)/2}}{C_{\nu, \nu+1}}
h^{(\nu)}(\vkappa) d \vkappa d \mu(U,V),
\label{eqn:chGOEJ}
\end{equation}
where $d \mu(U,V)$ is the normalized Haar measure of
the space ${\rm O}(N+\nu) \times {\rm O}(N)$ and
we have used the notation
$C_{\nu, \kappa}= 2^{N(N+2\nu-\kappa-1)/2} \pi^{-N/2} 
\prod_{i=1}^{N} \{ \Gamma (i/2)
\Gamma((i+2\nu+1-\kappa)/2)\}$
and thus $C_{\nu, \nu+1}=2^{N(N+\nu-2)/2} \pi^{-N/2}$
$\prod_{i=1}^{N} \{ \Gamma(i/2) \Gamma((i+\nu)/2) \}$.
The probability density of $\vkappa$
with (\ref{eqn:condition2}) is given as
\cite{VZ93,Ver94,JSV96,SV98}
$$
q_{\nu}^{\rm chGOE}(\vkappa; t) 
= \frac{t^{-N(N+\nu)/2}}{C_{\nu, \nu+1}}
\exp \left\{ - \frac{|\vkappa|^2}{2t} \right\} 
h^{(\nu)}(\vkappa).
$$

By setting $\kappa_{i}(t)=\sqrt{\lambda_{i}(t)}, 1 \leq i \leq N$
in (\ref{eqn:LW0}) with $\beta=1$,
we can show that $\vkappa(t)=(\kappa_{1}(t), \kappa_{2}(t),
\cdots, \kappa_{N}(t))$ solves 
(\ref{eqn:basic1}) with $(\beta, \gamma)=(1, \nu)$.
If we denote the transition probability density of this process
$\vkappa(t)$ by
$p^{(\nu)'}(s, \, \cdot \, ; t, \, \cdot \,)$ 
for $0 \leq s < t < \infty$, then
$p^{(\nu)'}(0, \0; t, \vkappa)
=q_{\nu}^{\rm chGOE}(\vkappa; t), \, t >0$.

\section{\large HERMITIAN MATRIX-VALUED PROCESSES WITH
ADDITIONAL SYMMETRIES}

\subsection{Subspaces of unitary and hermitian matrices}

The Pauli spin matrices are defined as
$$
\sigma_{1}=\left(\matrix{0 & 1 \cr 1 & 0}\right), \quad
\sigma_{2}=\left(\matrix{0 & -\sqrt{-1} \cr
\sqrt{-1} & 0} \right), \quad
\sigma_{3}=\left(\matrix{1 & 0 \cr 0 & -1} \right),
$$
which satisfy the algebra
$\sigma_{\mu}^2=I_{2},\mu=1,2,3$,
and
$\sigma_{\mu} \sigma_{\rho} = \sqrt{-1}
\sum_{\omega=1}^{3} \varepsilon_{\mu \rho \omega}
\sigma_{\omega}$
for $1 \leq \mu \not= \rho \leq 3$,
where $I_{N}$ denotes the $N \times N$ unit matrix and
$\varepsilon_{\mu \rho \omega}$ the totally
antisymmetric unit tensor.
They give the infinitesimal generators $\{X_{\mu}\}$ of
${\rm SU}(2)$ by $X_{\mu}=\sqrt{-1} \sigma_{\mu}/2$.
For $N \geq 2$, define the $2N \times 2N$ matrices
$\Sigma_{\mu}=I_{N} \otimes \sigma_{\mu}, 
\mu=1,2,3$.
The matrices $\{\Sigma_{\mu}\}$ satisfy the same
algebra as $\{\sigma_{\mu}\}$. 
We will use $\sigma_{0}$ to represent $I_{2}$.

We introduce six spaces of matrices as subspaces
of ${\cal H}(2N)$,
$$
{\cal H}_{\mu \pm}(2N) = \{ H \in {\cal H}(2N):
H^{T} \Sigma_{\mu} = \pm \Sigma_{\mu} H \}, \quad
\mu=1,2,3.
$$
It is easy to see that 
${\cal H}_{3+}(2N)={\cal S}(2N)$
and 
${\cal H}_{3-}(2N)=\sqrt{-1}{\cal A}(2N)$.
Since we have already studied the matrix-valued process
in ${\cal S}(N)$ as the example (ii) in Sec.II.B,
we will consider here the five subspaces of
${\cal H}(2N)$; $\sqrt{-1} {\cal A}(2N)$ and
$\{{\cal H}_{\mu \sigma}(2N)\}$ with
$\mu=1,2, \sigma=\pm$.
We also introduce the three subspaces of ${\rm U}(2N)$:
\begin{eqnarray}
{\rm U}_{0}(2N) &=& \{ U \in {\rm U}(2N):
U^{T}U=\Sigma_{1} \}, \nonumber\\
{\rm U}_{\mu} (2N) &=& \{ U \in {\rm U}(2N):
U^{T} \Sigma_{\mu} U = \Sigma_{\mu}^{T} \},
\quad \mu=1,2. \nonumber
\end{eqnarray}
The conditions imply that these subspaces,
${\cal H}_{\mu \sigma}(2N)$ and ${\rm U}_{\mu}(2N)$, have
additional symmetries compared to
${\cal H}(2N)$ and ${\rm U}(2N)$.
Concerning the eigenvalues and eigenvectors
of the hermitian matrices, the following lemma may be easily proved.

\begin{lem}
\label{thm:subspaces}
Assume that $\Omega$ denotes a diagonal matrix
in the form ${\rm diag} 
\{\omega_{1}, \omega_{2}, \cdots, \omega_{N}\}$ with
$\omega_{1} \leq \omega_{2} \leq \cdots \leq \omega_{N}$.
\begin{description}
\item{(i)} \quad
Any $H \in \sqrt{-1} {\cal A}(2N)$ can be diagonalized
by $U \in {\rm U}_{0}(2N)$ as
$U^{\dagger} H U= \Omega \otimes \sigma_{3}$.
\item{(ii)} \quad
For $\mu=1,2$ 
any $H \in {\cal H}_{\mu +}(2N)$ can be diagonalized
by $U \in {\rm U}_{\mu}(2N)$ as
$U^{\dagger} H U= \Omega \otimes \sigma_{0}$.
\item{(iii)} \quad
For $\mu=1,2$
any $H \in {\cal H}_{\mu -}(2N)$ can be diagonalized
by $U \in {\rm U}_{\mu}(2N)$ as
$U^{\dagger} H U= \Omega \otimes \sigma_{3}$.
\end{description}
\end{lem}

\vskip 0.5cm
\noindent{\bf Remark} 
\begin{description}
\item{(a)} \quad 
Observing the pairing of eigenvalues in a way,
$(\omega_{i}, - \omega_{i}),
1 \leq i \leq N$, for $\sqrt{-1}{\cal A}(2N)$
stated in Lemma \ref{thm:subspaces} (i),
the Gaussian random-matrix ensemble of antisymmetric hermitian
matrices was discussed by Mehta in Section 3.4 of \cite{Meh91}.

\item{(b)} \quad
The condition for ${\rm U}_{2}(2N)$ addition to the unitarity 
is equivalent with
$J=U J U^{T}$,
where
$$
   J=I_{N} \otimes \left( \matrix{0 & 1 \cr -1 & 0} \right).
$$
Then ${\rm U}_{2}(2N)$ forms the $N$-dimensional symplectic
group. That is, 
${\rm U}_{2}(2N) = {\rm Sp}(N, \C) \cap {\rm U}(2N)$.
(It is called the unitary-symplectic group
${\rm USp}(2N)$ in \cite{Gil74}.)
The matrices $H \in {\cal H}_{2+}(2N)$ are said to be
{\it self-dual hermitian matrices} 
in the random matrix theory \cite{Meh91}.
The pairwise degeneracy stated in Lemma 
\ref{thm:subspaces}
(ii) for ${\cal H}_{2+}(2N)$ is known as the {\it Kramers doublet}
in the quantum mechanics.

\item{(c)} \quad 
The condition for ${\cal H}_{2-}(2N)$ addition to hermiticity
is rewritten as
$
  H^{T} J + J H=0,
$
which means that $H \in {\cal H}_{2-}(2N)$ satisfies
the symplectic Lie algebra 
(see for example \cite{FH91}), that is,
${\cal H}_{2-}(2N) = \mathfrak{sp}(2N, \C) \cap {\cal H}(2N)$.
Similarly, we can see 
${\cal H}_{1-}(2N) = \mathfrak{so}(2N, \C) \cap {\cal H}(2N)$,
where $\mathfrak{so}(2N, \C)$ denotes the orthogonal Lie algebra.
We can also see that 
${\rm U}_{1}(2N) = {\rm SO}(2N, \C) \cap {\rm U}(2N)$,
where ${\rm SO}(2N, \C)$ denotes the orthogonal Lie group.

\item{(d)} \quad 
We can see that 
${\cal H}_{\mu-}(2N) \cong \widehat{\cal H}_{\mu -}(2N),
\mu=1,2,$ where
\begin{eqnarray}
\widehat{\cal H}_{1-}(2N) &=& \left\{
H =\left( \matrix{ H_1 & A_2 \cr A_2^{\dagger} & -H_1^{T} } \right): 
\; H_1\in {\cal H}(N), A_{2} \in {\cal A}(N; \C) \right\}, 
\nonumber\\
\widehat{\cal H}_{2-}(2N) &=& \left\{
H =\left( \matrix{ H_1 & A_2 \cr A_2^{\dagger} & -H_1^{T} } \right): 
\; H_1\in {\cal H}(N), A_{2} \in {\cal S}(N; \C) \right\},
\nonumber
\end{eqnarray}
where ${\cal S}(N; \C)$ and ${\cal A}(N; \C)$ denote
the spaces of the $N \times N$ complex symmetric and 
complex antisymmetric matrices, respectively.
Altland and Zirnbauer studied $\widehat{\cal H}_{2-}(2N)$
and $\widehat{\cal H}_{1-}(2N)$
as the sets of the Hamiltonians in the Bogoliubov-de Gennes
formalism for the BCS mean-field theory
of superconductivity,
where the pairing of positive and negative eigenvalues
$(\omega_{i}, - \omega_{i}), 1 \leq i \leq N$,
stated in Lemma \ref{thm:subspaces} (iii) for
$\mu=1$ and 2 represents the particle-hole
symmetry in the Bogoliubov-de Gennes theory.
They called $\widehat{\cal H}_{2-}(2N)$
and $\widehat{\cal H}_{1-}(2N)$ the sets of hermitian matrices
in the symmetry classes C and D \cite{Zir96,AZ96,AZ97}, 
since $\mathfrak{sp}(2N, \C)=\mbox{C}_{N}$ and
$\mathfrak{so}(2N, \C)=\mbox{D}_{N}$ in Cartan's notations
(see \cite{Hel78}).
\end{description}

\subsection{Representation using Pauli matrices
and application of Bru's theorem}

Let $B_{ij}^{\rho}(t)$, $\widetilde{B}^{\rho}_{ij}(t)$, 
$0 \leq \rho \leq 3$, $1\le i \leq j \leq N$
be independent one-dimensional standard Brownian motions
starting from the origin. 
Put
\begin{equation}
s_{ij}^{\rho}(t)
=
\left\{
   \begin{array}{ll}
      \displaystyle{
      \frac{1}{\sqrt{2}} B_{ij}^{\rho}(t),
      } & 
   \mbox{if} \ i < j, \\
        & \\
        B_{ii}^{\rho}(t), & \mbox{if} \ i=j,\\
   \end{array}\right. 
   \quad {\rm and } \quad
a_{ij}^{\rho}(t)
=
\left\{
   \begin{array}{ll}
      \displaystyle{
      \frac{1}{\sqrt{2}} \widetilde{B}_{ij}^{\rho}(t),
      } & 
   \mbox{if} \ i < j, \\
        & \\
    0, & \mbox{if} \ i=j, \\
   \end{array}\right. 
   \label{eqn:sar}
\end{equation} 
with $s_{ij}^{\rho}(t)=s_{ji}^{\rho}(t)$ and
$a_{ij}^{\rho}(t)=-a_{ji}^{\rho}(t)$ for
$i > j$
and define 
$s^{\rho}(t)=(s^{\rho}_{ij}(t))_{1 \leq i, j \leq N}
\in {\cal S}(N), \, t \in [0, \infty)$
and
$a^{\rho}(t)=(a^{\rho}_{ij}(t))_{1 \leq i, j \leq N}
\in {\cal A}(N), t \in [0, \infty)$,
for $0 \leq \rho \leq 3$.

We can see that the hermitian matrix-valued process
given as the first example (i)
in Sec. II.B can be represented, if we double
the size of matrix to $2N$, as
$\Xi(t)=\sum_{\rho=0}^{3} \{
(s^{\rho}(t) \otimes \sigma_{\rho})
+ \sqrt{-1}(a^{\rho}(t) \otimes \sigma_{\rho}) \}$.
By choosing four terms in the eight terms,
we define the following four different types of $2N \times 2N$
hermitian matrix-valued processes:
$$
\Xi_{\mu \sigma}(t)=\sum_{\rho=0}^{3}
( \xi_{\mu \sigma}^{\rho}(t) \otimes \sigma_{\rho} ) 
\in {\cal H}_{\mu \sigma}(2N) \quad \mbox{for} \quad
\mu=1,2, \quad \sigma=\pm,
$$
where 
\begin{eqnarray}
&& 
\xi_{\mu +}^{\rho}(t) =
\left\{
   \begin{array}{ll}
      s^{\rho}(t) & 
  \mbox{if} \ \mu = 1, \quad \rho \not= 3 \quad 
  \mbox{or} \quad \mu=2, \quad \rho=0, \\
  & \\
      \sqrt{-1} a^{\rho}(t) 
      & \mbox{if} \ \mu = 1, \quad \rho = 3 \quad
      \mbox{or} \quad \mu=2, \quad \rho \not= 0,
   \end{array}\right.
\nonumber\\
&& 
\xi_{\mu -}^{\rho}(t) =
\left\{
   \begin{array}{ll}
      \sqrt{-1} a^{\rho}(t) & 
  \mbox{if} \ \mu = 1, \quad \rho \not= 3 \quad 
  \mbox{or} \quad \mu=2, \quad \rho=0, \\
  & \\
      s^{\rho}(t) 
      & \mbox{if} \ \mu = 1, \quad \rho = 3 \quad \mbox{or} 
\quad \mu=2, \quad \rho \not= 0.
   \end{array}\right.
\nonumber
\end{eqnarray}

We apply Theorem \ref{thm:BruThm} to the
five processes, $\sqrt{-1} {\cal A}(2N)$ and
$\{\Xi_{\mu \sigma}(t)\}$ with
$\mu=1,2, \sigma=\pm$.
The results are listed below.

\begin{description}
\item{(a)} $\sqrt{-1}{\cal A}(2N)$:
Since
$ \Gamma_{ij}(t) =  
\{ 1-((\Sigma_{1})_{ij})^2 \}/2,
\, 1 \leq i, j \leq 2N$,
the equations of nonnegative eigenvalues are
$$
d \omega_{i}(t) = \frac{1}{\sqrt{2}}dB_{i}(t) + \frac{1}{2}
\sum_{j: 1 \leq j \leq N, j \not=i} \left\{
\frac{1}{\omega_{i}(t)-\omega_{j}(t)}
+ \frac{1}{\omega_{i}(t)+\omega_{j}(t)} \right\} dt, \quad
1 \leq i \leq N.
$$
By changing the time unit as $t \to 2t$, this equation 
can be identified with (\ref{eqn:basic1}) with
$(\beta, \gamma)=(2, 0)$.

\item{(b)} ${\cal H}_{1+}(2N)$:
Since 
$\Gamma_{ij}(t)=\left\{ 1+((\Sigma_{1})_{ij})^2 \right\},
\, 1 \leq i, j \leq 2N$,
the distinct eigenvalues solve Dyson's Brownian motion model
(\ref{eqn:Dyson1}) with $\beta=4$.

\item{(c)} ${\cal H}_{1-}(2N)$:
We see
$\Gamma_{ij}(t)=\left\{ 1-((\Sigma_{1})_{ij})^2 \right\},
\, 1 \leq i, j \leq 2N$. Then
the nonnegative eigenvalues solve
the equations (\ref{eqn:basic1}) 
with $(\beta, \gamma)=(2, 0)$.

\item{(d)} ${\cal H}_{2+}(2N)$:
Since $\Gamma_{ij}(t)=\left\{ 1+((\Sigma_{2})_{ij})^2 \right\},
\, 1 \leq i, j \leq 2N$,
the distinct eigenvalues solve the equations
(\ref{eqn:Dyson1}) with $\beta=4$

\item{(e)} ${\cal H}_{2-}(2N)$:
We can see 
$\Gamma_{ij}(t)=\left\{ 1-((\Sigma_{2})_{ij})^2 \right\},
\, 1 \leq i, j \leq 2N$. Then
the nonnegative eigenvalues solve the equation
(\ref{eqn:basic1}) with $(\beta, \gamma)=(2, 1)$.
\end{description}

\subsection{Relation with standard and 
nonstandard random matrix theories}

\vskip 0.5cm
\noindent{(i)} \quad
The eigenvalues of any matrix in the space
${\cal H}_{2+}(2N) \cong \R^{d[{\rm A}'']}$
with $d[{\rm A}'']=N(2N-1)$ 
are pairwise degenerated (the Kramers doublets) 
as $\vlambda=(\omega_{1}, \omega_{1}, \omega_{2},
\omega_{2}, $
$\cdots, \omega_{N}, \omega_{N})$.
We assume that the $N$ distinct eigenvalues are
always arranged in the increasing order 
$\omega_{1} \leq \omega_{2} \leq \cdots \leq \omega_{N}$.
For GSE with variance $t$,
the probability density of the $N$ distinct eigenvalues 
in this ordering is given by \cite{Meh91}
$$
q^{\rm GSE}(\vomega; t)
= \frac{t^{-d[{\rm A}'']/2}}{C[{\rm A}'']} 
\exp \left\{ - \frac{|\vomega|^2}{2t} \right\} 
h^{\rm A}(\vomega)^4,
$$
where $C[{\rm A}'']=(2\pi)^{N/2} \prod_{i=1}^{N} \Gamma(2i)$.
If we denote the transition probability density of the process
(\ref{eqn:Dyson1}) with $\beta=4$ by
$p^{\rm A''}(s, \, \cdot \, ; t, \, \cdot \,)$ 
for $0 \leq s < t < \infty$, then
$p^{\rm A''}(0, \0; t, \vomega)
=q^{\rm GSE}(\vomega; t), \, t >0$.

\vskip 0.5cm
\noindent{(ii)} \quad
We can see that ${\cal H}_{2-}(2N) \cong \R^{d[{\rm C}]}$
and ${\cal H}_{1-}(2N) \cong \R^{d[{\rm D}]}$ with
$d[{\rm C}]=N(2N+1)$ and $d[{\rm D}]=N(2N-1)$.
The probability densities of the processes
$\Xi_{2-}(t)$ and $\Xi_{1-}(t)$ with respect to the volume elements
${\cal V}(d H)$ of ${\cal H}_{2-}(2N)$ 
and ${\cal V}'(d H)$ of ${\cal H}_{1-}(2N)$ are given by
$$
\mu^{\rm C}
(H; t)= \frac{t^{-d[{\rm C}]/2}}{c[{\rm C}]}
\exp \left\{ - \frac{1}{4t} {\rm Tr} \, H^2 \right\},
\quad 
\mu^{\rm D}
(H; t)= \frac{t^{-d[{\rm D}]/2}}{c[{\rm D}]}
\exp \left\{ - \frac{1}{4t} {\rm Tr} \, H^2 \right\},
$$
where $c[{\rm C}]=2^{3N/2} \pi^{N(2N+1)/2}$ and
$c[{\rm D}]=2^{N/2} \pi^{N(2N-1)/2}$, respectively.
As stated in Lemma \ref{thm:subspaces} (iii), 
the eigenvalues are in the form
$\vlambda(t)=(\omega_{1}(t), -\omega_{1}(t), \omega_{2}(t),$
$-\omega_{2}(t), \cdots, \omega_{N}(t), -\omega_{N}(t))$.
We will assume that
\begin{equation}
0 \leq \omega_{1} \leq \omega_{2} \leq \cdots \leq \omega_{N}.
\label{eqn:condition4}
\end{equation}
Then we have the expressions for volume elements
\begin{equation}
{\cal V}(d H)=\frac{c[{\rm C}]}{C[{\rm C}]} h^{\rm C}(\vomega)^2
d \vomega dU,
\quad
{\cal V'}(d H)=\frac{c[{\rm D}]}{C[{\rm D}]} h^{\rm D}(\vomega)^2
d \vomega dU', 
\label{eqn:CDJ}
\end{equation}
where $dU$ and $dU'$ denote the Haar measures of
${\rm U}_{2}(2N)$ and ${\rm U}_{1}(2N)$, respectively, normalized as\\
$\int_{{\rm U}_{2}(2N)}d U=1$ and
$\int_{{\rm U}_{1}(2N)} dU'=1$. 
Here $C[{\rm C}]=C_{1/2}
=(\pi/2)^{N/2} \prod_{i=1}^{N} \Gamma(2i)$ and
$C[{\rm D}]=C_{-1/2}=(\pi/2)^{N/2} \prod_{i=1}^{N} \Gamma(2i-1)$, 
and
$h^{\rm C}(\vomega) \equiv h^{(1)}(\vomega),
h^{\rm D}(\vomega) \equiv h^{(0)}(\vomega)$.
At each time $t >0$, 
for any $U \in {\rm U}_{2}(2N)$,
the probability
$\mu^{\rm C}(H; t) {\cal V}(d H)$ is
invariant under the automorphism
$H \to U^{\dagger} H U$ for $H \in {\cal H}_{2-}(2N)$, 
and for any $U' \in {\rm U}_{1}(2N)$,
$\mu^{\rm D}(H; t) {\cal V'}(d H)$ is
invariant under the automorphism
$H \to U'^{\dagger} H U'$ for $H \in {\cal H}_{1-}(2N)$.
Altland and Zirnbauer named
these two Gaussian random-matrix ensembles
the classes C and D, respectively 
(see Remark (d) in Sec.III.A) \cite{AZ96,AZ97,Zir96}.
The probability densities
of the $N$ nonnegative eigenvalues 
with the condition (\ref{eqn:condition4})
are then obtained as
$$
q^{\sharp}(\vomega; t)
= \frac{t^{-d[\sharp]/2}}{C[\sharp]} 
\exp \left\{ - \frac{|\vomega|^2}{2t} \right\} 
h^{\sharp}(\vomega)^2 \quad
\mbox{for} \quad \sharp=\mbox{C, D}.
$$
If we denote the transition probability densities of the processes
(\ref{eqn:basic1}) with $(\beta, \gamma)=(2,1)$
and with $(\beta, \gamma)=(2,0)$ by
$p^{\rm C}(s, \, \cdot \, ; t, \, \cdot \,)$ and
$p^{\rm D}(s, \, \cdot \, ; t, \, \cdot \,)$ 
for $0 \leq s < t < \infty$, respectively, then
\begin{equation}
p^{\sharp}(0, \0; t, \vomega)
=q^{\sharp}(\vomega; t),
\quad t >0 \quad
 \mbox{for} \quad \sharp=\mbox{C, D}.
\label{eqn:CD1}
\end{equation}

\subsection{Real symmetric matrix-valued processes}

Here after, we denote the hermitian matrix-valued processes
$\Xi_{2-}(t)$ and $\Xi_{1-}(t)$ by 
$\Xi^{\rm C}(t)$ and $\Xi^{\rm D}(t)$, respectively.
They are given by
\begin{eqnarray}
&& \Xi^{\rm C}(t)= \sqrt{-1} a^{0}(t) \otimes \sigma_{0}
+ s^{1}(t) \otimes \sigma_{1}
+ s^{2}(t) \otimes \sigma_{2}
+ s^{3}(t) \otimes \sigma_{3}, \nonumber\\
&& \Xi^{\rm D}(t)= \sqrt{-1} a^{0}(t) \otimes \sigma_{0}
+ \sqrt{-1} a^{1}(t) \otimes \sigma_{1}
+ \sqrt{-1} a^{2}(t) \otimes \sigma_{2}
+ s^{3}(t) \otimes \sigma_{3}.
\label{eqn:XiCD}
\end{eqnarray}
Since $\sigma_{\rho}, \rho=0,1,3$, are real matrices and
$\sigma_{2}$ is a pure imaginary matrix, if we
define the processes as
\begin{equation}
\Xi^{\rm C'}(t)= s^{1}(t) \otimes \sigma_{1}
+ s^{3}(t) \otimes \sigma_{3}, \quad
\Xi^{\rm D'}(t)= 
\sqrt{-1} a^{2}(t) \otimes \sigma_{2}
+ s^{3}(t) \otimes \sigma_{3}, 
\label{eqn:XiCDp1}
\end{equation}
then 
$\Xi^{\rm C'}(t) \in {\cal S}_{2-}(2N)$ and
$\Xi^{\rm D'}(t) \in {\cal S}_{1-}(2N)$, where
${\cal S}_{2 -}(2N) \equiv \{S \in {\cal S}(2N):
S^{T} \Sigma_{2}=-\Sigma_{2} S \} 
\cong \R^{d[{\rm C}']}$ and
${\cal S}_{1 -}(2N) \equiv \{S \in {\cal S}(2N):
S^{T} \Sigma_{1}=-\Sigma_{1} S \} 
\cong \R^{d[{\rm D}']}$ with
$d[{\rm C}']=N(N+1)$ and $d[{\rm D}']=N^2$.
The probability densities of
$\Xi^{\rm C'}(t)$ and $\Xi^{\rm D'}(t)$ are given by
$$
\mu^{\rm C'}
(S; t)= \frac{t^{-d[{\rm C}']/2}}{c[{\rm C}']}
\exp \left\{ - \frac{1}{4t} {\rm Tr} \, S^2 \right\},
\quad 
\mu^{\rm D'}
(S; t)= \frac{t^{-d[{\rm D}']/2}}{c[{\rm D}']}
\exp \left\{ - \frac{1}{4t} {\rm Tr} \, S^2 \right\}
$$
with
$c[{\rm C}']=2^{N} \pi^{N(N+1)/2}$ and
$c[{\rm D}']=2^{N/2} \pi^{N^2/2}$, respectively.
Set ${\rm O}_{2}(2N)={\rm O}(2N) \cap {\rm Sp}(2N;\R)$
and
${\rm O}_{1}(2N)={\rm O}(2N) \cap {\rm SO}(2N;\R)$
and denote their normalized Haar measures by $dV$ and $dV'$,
respectively
The eigenvalues are in the form
$\vlambda(t)=(\omega_{1}(t), -\omega_{1}(t), \omega_{2}(t),
-\omega_{2}(t)$,
$ \cdots, \omega_{N}(t), -\omega_{N}(t))$.
Under the condition (\ref{eqn:condition4}),
we have the expressions for volume elements
\begin{equation}
{\cal V}(d S)=\frac{c[{\rm C}']}{C[{\rm C}']} h^{\rm C}(\vomega)
d \vomega dV, \quad 
{\cal V'}(d S)=\frac{c[{\rm D}']}{C[{\rm D}']} h^{\rm D}(\vomega)
d \vomega dV', 
\label{eqn:CDpJ}
\end{equation}
where $C[{\rm C}']=C_{1/2,1}
=\prod_{i=1}^{N} \Gamma(i)$ and
$C[{\rm D}']=C_{-1/2,0}
=2^{(N-2)/2}
\Gamma(N/2) \prod_{i=1}^{N-1} \Gamma(i)$.
The probability densities
of the $N$ nonnegative eigenvalues 
with (\ref{eqn:condition4})
are given as
$$
q^{\sharp'}(\vomega; t)
= \frac{t^{-d[\sharp']/2}}{C[\sharp']} 
\exp \left\{ - \frac{|\vomega|^2}{2t} \right\} 
h^{\sharp}(\vomega) 
\quad \mbox{for} \quad \sharp=\mbox{C, D}.
$$
It is remarked that the random-matrix
ensemble with the distributions $\mu^{\rm C'}(S; t)$,
whose nonnegative eigenvalue distribution is given by
$q^{\rm C'}(\vomega; t)$, is 
the random-matrix ensemble in the symmetry class CI 
studied by Altland and Zirnbauer \cite{AZ96,AZ97,Zir96}.

By applying Theorem \ref{thm:BruThm}, we can show that
the nonnegative eigenvalues of $\Xi^{\rm C'}(t)$
solve the equations (\ref{eqn:basic1}) with
$(\beta, \gamma)=(1,1)$ and those of $\Xi^{\rm D'}(t)$
the equations (\ref{eqn:basic1}) with
$(\beta,\gamma)=(1,0)$.
If we denote the transition probability densities of these
processes by
$p^{\rm C'}(s, \, \cdot \, ; t, \, \cdot \,)$ and
$p^{\rm D'}(s, \, \cdot \, ; t, \, \cdot \,)$ 
for $0 \leq s < t < \infty$, respectively, then
$p^{\sharp'}(0, \0; t, \vomega)
=q^{\sharp'}(\vomega; t), \, t >0$
for $\sharp=\mbox{C, D}$.

\section{\large TEMPORALLY HOMOGENEOUS PROCESSES}

Assume that $\nu > -1$, and we consider the process
$\Y^{(\nu)}(t)=(Y_{1}^{(\nu)}(t), Y_{2}^{(\nu)}(t),
\cdots, Y_{N}^{(\nu)}(t)), t \in [0, \infty)$,
which solves the stochastic differential equations
(\ref{eqn:basic1}) with $(\beta, \gamma)=(2, (2\nu+1)/2)$,
that is, 
\begin{equation}
d Y^{(\nu)}_{i}(t)=d B_{i}(t)+
\left[
\frac{2\nu+1}{2 Y^{(\nu)}_{i}(t)}
+ \sum_{j: j \not=i} 
\left\{ \frac{1}{Y^{(\nu)}_{i}(t)-Y^{(\nu)}_{j}(t)}
+\frac{1}{Y^{(\nu)}_{i}(t)+Y^{(\nu)}_{j}(t)} \right\} \right] dt,
\label{eqn:Ynu1}
\end{equation}
$1 \leq i \leq N$.
Remark that if $\nu=1/2$ and $-1/2$,
the equation is reduced to (\ref{eqn:basic1})
with $(\beta, \gamma)=(2,1)$ and $(\beta, \gamma)=(2,0)$,
respectively.
The Kolmogorov backward equation (the Fokker-Planck equation)
for (\ref{eqn:Ynu1}) is
$$
\frac{\partial}{\partial t} p^{(\nu)}(s, \x; t, \y)
= \frac{1}{2} \Delta_{\x} 
p^{(\nu)}(s, \x; t, \y)
+ {\bf b}({\bf x}) \cdot \nabla_{\x} 
p^{(\nu)}(s, \x; t, \y),
$$
where ${\bf b}({\bf x})=(b_{1}({\bf x}), 
\cdots, b_{N}({\bf x}))$ with
$b_{i}(\x)=(\partial / \partial x_{i})
\ln h^{((2\nu+1)/2)}(\x)$.
By simple calculation, we can confirm the following.

\begin{lem}
\label{thm:Bessel1}
Set
\begin{equation}
f^{(\nu)}(t, \y|\x) = \det_{1 \leq i, j \leq N}
\Bigg[ G^{(\nu)}(t, y_{j}|x_{i}) \Bigg].
\label{eqn:fNnu}
\end{equation}
Then the transition probability density $p^{(\nu)}(s, \x; t, \y)$
from the state $\x$ at time $s$ to the state
$\y$ at time $t (> s)$ of the process (\ref{eqn:Ynu1}) is given by
\begin{equation}
p^{(\nu)}(s, \x; t, \y)=
\frac{1}{h^{(0)}(\x)}
f^{(\nu)}(t-s, \y|\x) h^{(0)}(\y), 
\quad \x, \y \in \W_{N}^{\rm C}.
\label{eqn:Bessel2}
\end{equation}
\end{lem}

Since 
$I_{1/2}(x)=(e^{x}-e^{-x})/\sqrt{2 \pi x}, \,
I_{-1/2}(x)=(e^{x}+e^{-x})/\sqrt{2 \pi x}$, 
if we set
\begin{eqnarray}
\label{eqn:GC}
G^{\rm C}(t, y|x)= \frac{e^{-(y-x)^2/2t}-e^{-(y+x)^2/2t}}
{\sqrt{2 \pi t}},
\quad
G^{\rm D}(t, y|x)= \frac{e^{-(y-x)^2/2t}+e^{-(y+x)^2/2t}}
{\sqrt{2 \pi t}},
\label{eqn:GCD}
\end{eqnarray}
and
$f^{\sharp}(t, \y|\x) = \det_{1 \leq i, j \leq N}
[ G^{\sharp}(t, y_{j}|x_{i})], \,
\sharp=\mbox{C, D}$,
then
\begin{equation}
p^{(1/2)}(s, \x; t, \y)
= \frac{f^{\rm C}(t-s, \y|\x) h^{\rm C}(\y)}{h^{\rm C}(\x)}, \quad
p^{(-1/2)}(s, \x; t, \y)
= \frac{f^{\rm D}(t-s, \y|\x)h^{\rm D}(\y)}{h^{\rm D}(\x)}.
\label{eqn:pNCD}
\end{equation}
The above implies the following.
Let
$\W_{N}^{\rm C}= \{ \x \in \R^{N}:
0 < x_{1} < x_{2} < \cdots < x_{N} \}$ and
$\W_{N}^{\rm D}=\{ \x \in \R^{N}:
|x_{1}| < x_{2} < \cdots < x_{N} \}$.
The former is the Weyl chamber of type $\mbox{C}_{N}$
and the latter of type $\mbox{D}_{N}$ \cite{FH91}.
Since $h^{\rm C}$ and $h^{\rm D}$ vanish at the boundaries
of the Weyl chambers $\W_{N}^{\rm C}$ and $\W_{N}^{\rm D}$,
respectively, 
(\ref{eqn:pNCD}) implies that the processes 
$\Y^{(1/2)}(t)$ and $\Y^{(-1/2)}(t)$ can be regarded as
the $N$-dimensional absorbing Brownian motions in $\W_{N}^{\rm C}$ 
and in $\W_{N}^{\rm D}$, respectively.
That is, if $\Y^{(1/2)}(0) \in \W_{N}^{\rm C}$ 
and $\Y^{(-1/2)}(0) \in \W_{N}^{\rm D}$,
then $\Y^{(1/2)}(t) \in \W_{N}^{\rm C}$ and
$\Y^{(-1/2)}(t) \in \W_{N}^{\rm D}$
for all $t >0$ with probability 1.
Moreover, we notice that (\ref{eqn:GCD}) are
the heat-kernels of the one-dimensional Brownian
motion with an absorbing wall at the origin,
and of the one-dimensional reflecting Brownian motion,
respectively \cite{RY98}. Then, we can also interpret the process
$\Y^{(1/2)}(t)$ as the $N$-particle system
of Brownian motions conditioned never to collide with
each other nor with the wall at the origin in
one-dimension \cite{KTNK03}, and
the process $\Y^{(-1/2)}(t)$ as
the $N$-particle system of reflecting Brownian motions
conditioned never to collide with each other.
For $\sharp=$C and D, define
\begin{equation}
{\cal N}^{\sharp}(t, \x)=
\int_{\W_{N}^{\sharp}} d \y \,
f^{\sharp}(t, \y|\x),
\quad \x \in \W_{N}^{\sharp}.
\label{eqn:NNCD}
\end{equation}
${\cal N}^{\rm C}(t, \x)$ is the probability that 
$N$ Brownian motions starting from $\x \in \W_{N}^{\rm C}$
does not collide with
each other nor with the wall at the origin up to time $t$,
and ${\cal N}^{\rm D}(t, \x)$ is equal to the probability that 
$N$ reflecting Brownian motions starting from $\x \in \W_{N}^{\rm D}$
does not collide with
each other up to time $t$, respectively.
We will show their long-time asymptotics in the next section.
We can prove the following, which are consistent with
(\ref{eqn:Laguerre6}) and (\ref{eqn:CD1}).

\begin{lem}
\label{thm:xzero}
For $\nu > -1$ with fixed $t \in (0, \infty)$,
assume $\y \in \W_{N}^{\rm C}$. Then
\begin{equation}
\lim_{|\x| \to 0} p^{(\nu)}(0, \x; t, \y)
= \frac{t^{-N(N+\nu)}}{C_{\nu}} 
\exp \left\{ - \frac{|\y|^2}{2t} \right\}
h^{((2\nu+1)/2)}(\y)^2.
\label{eqn:limit0}
\end{equation}
In particular, if $\nu \in \N$,
\begin{equation}
\lim_{|\x|\to 0} 
p^{(\nu)}(0, \x; t, \y)
= q_{\nu}^{\rm chGUE}(\y; t),
\label{eqn:limit1}
\end{equation}
and
\begin{equation}
\lim_{|\x|\to 0} 
p^{(1/2)}(0, \x; t, \y)
= q^{\rm C}(\y; t), \qquad
\lim_{|\x|\to 0} 
p^{(-1/2)}(0, \x; t, \y)
= q^{\rm D}(\y; t).
\label{eqn:limit2}
\end{equation}
\end{lem}
\noindent{\it Proof.} \quad
By definition (\ref{eqn:fNnu}) with (\ref{eqn:Bessel1}),
if $x_{i}>0, \, 1 \leq \forall i \leq N$,
$f^{(\nu)}(t, \y|\x) =$  
$(1/t^{N})\prod_{k=1}^{N}
(y_{k}^{\nu+1}/x_{k}^{\nu})$
$e^{-(|\x|^2+|\y|^2)/2t}
\det_{1 \leq i, j \leq N} [
I_{\nu} (x_{i} y_{j}/t) ]$.
We can use (\ref{eqn:Schur2}) in Appendix A
by changing the variables $x_{i} \to x_{i}^{2}/2t$
and $y_{j} \to y_{j}^2/2t$ 
to evaluate $\det_{1 \leq i, j \leq N}
[I_{\nu}(x_{i}y_{j}/t)]$ and
obtain the asymptotic form of $f^{(\nu)}(t, \y|\x)$,
\begin{eqnarray}
f^{(\nu)}(t, \y|\x) &=& \frac{t^{-N(N+2\nu+1)/2}}{C_{\nu}} 
\prod_{1 \leq i < j \leq N} \left\{
\left(\frac{x_{j}}{\sqrt{t}}\right)^2
-\left(\frac{x_{i}}{\sqrt{t}}\right)^2 \right\} 
\nonumber\\
&& \times \prod_{1 \leq k < \ell \leq N}
(y_{\ell}^2-y_{k}^2) \prod_{m=1}^{N} y_{m}^{2\nu+1}
\exp \left\{ - \frac{|\y|^2}{2t} \right\}
\times 
\left( 1+ {\cal O} \left( \frac{|\x|}{\sqrt{t}} \right) \right)
\label{eqn:asymfNnu}
\end{eqnarray}
in $|\x|/\sqrt{t} \to 0$.
Using this form in (\ref{eqn:Bessel2}),
the limit (\ref{eqn:limit0}) is proved.
\qed

\section{\large TEMPORALLY INHOMOGENEOUS PROCESSES}

\subsection{Star topology}

\begin{figure}
\includegraphics[width=.5\linewidth]{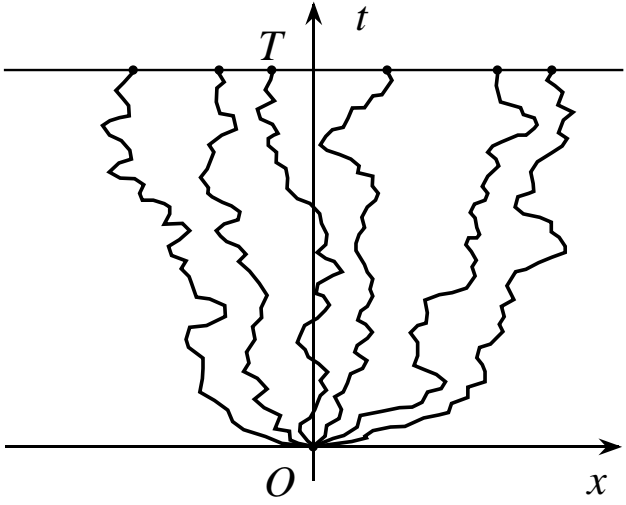}
\caption{Process $\X(t), t \in [0, T]$, with $\X(0)=\0$
showing star topology. \label{fig:star}}
\end{figure}

Using (\ref{eqn:fNA}) the probability that the Brownian motion
started at $\x \in \W_{N}^{\rm A}$ does not hit the boundary
of $\W_{N}^{\rm A}$ up to time $t > 0$ is given by
${\cal N}^{\rm A}(t, \x) = \int_{\W_{N}^{\rm A}}d \y \,
f^{\rm A}(t, \y|\x)$.
In the previous papers \cite{KT02,KT03a}, 
we gave the asymptotic form
\begin{equation}
f^{\rm A}(t, \y|\x) = \frac{t^{-N(N+1)/4}}{C[{\rm A}]}
h^{\rm A} \left( \frac{\x}{\sqrt{t}} \right)
h^{\rm A}(\y)
\exp \left\{-\frac{|\y|^2}{2t} \right\}
\times \left( 1+ {\cal O} \left( 
\frac{|\x|}{\sqrt{t}} \right) \right)
\label{eqn:asymfA}
\end{equation}
in $|\x|/\sqrt{t} \to 0$ and
showed that
${\cal N}^{\rm A}(t, \x) = (C[{\rm A}']/C[{\rm A}])
h^{\rm A} (\x/\sqrt{t})
\times (1+ {\cal O} ( |\x|/\sqrt{t}))$
as $|\x|/\sqrt{t} \to 0$.
This estimate gives that for $\x \in \W_{N}^{\rm A}$
the noncolliding probability decays in the power-law 
as $t \to \infty$ \cite{Fis84,Gra99,KGV00};
${\cal N}^{\rm A}(t, \x) \sim t^{-\psi[{\rm A}]}$
with the exponent $\psi[{\rm A}]=N(N-1)/4$.
(Note that (\ref{eqn:asymfA}) is derived readily by using 
(\ref{eqn:Schur1}) in Appendix A.)
For a given $T>0$, we defined
$$
g_{T}^{\rm A}(s,\x; t, \y )
= \frac{f^{\rm A}(t-s, \y|\x)
{\cal N}^{\rm A} (T-t,\y)}{{\cal N}^{\rm A} (T-s,\x)}
$$
for $ 0 \leq s < t \le T,\; \x, \y \in \W_{N}^{\rm A}$.
Using (\ref{eqn:asymfA}) we showed that
as $|\x|\to 0$ it converges to 
$g_{T}^{\rm A}(0, \0; t, \y)
=(T^{\psi[{\rm A}]} t^{-d[{\rm A}]/2}/C[{\rm A}'])
e^{-|\y|^2/2t}
h^{\rm A}(\y){\cal N}^{\rm A} (T-t,\y)$.
This function $g_{T}^{\rm A}(s,\x; t, \y)$
can be regarded as the transition 
probability density from the
state $\x \in \W_{N}^{\rm A} $ at time $s$ 
to the state $\y \in \W_{N}^{\rm A}$ at time $t (>s)$
conditioned to stay inside $\W_{N}^{\rm A}$ up to time $T$
and defines a temporally inhomogeneous diffusion process,
which we denoted by
$\X(t)=(X_1(t),X_2(t),\dots,X_N(t)), \, t\in [0,T]$ in Sec.I.
This represents the $N$-particle system of 
Brownian motions conditioned not 
to collide with each other in a finite time-interval $(0,T]$.
The process $\X(t), t \in [0, T]$, starting from
$\X(0)=\0$ is illustrated by Figure 1,
whose spatial-temporal path-configuration is said to be in
{\it star topology} in the theory of 
directed polymer networks \cite{EG95}.
As mentioned in Sec.I, this process exhibits
a transition of the eigenvalue statistics from GUE to GOE
\cite{KT02,KT03a}.

In the present section, we consider the temporally
inhomogeneous diffusion process associated with
$\Y^{(\nu)}(t)$ studied in the previous section.
We consider the $N$-particle system of generalized meanders
(\ref{eqn:meander})
conditioned that they never collide with each other 
for a time interval $[0, T]$. 
The transition probability density is given by
\begin{equation}
g_{T}^{(\nu, \kappa)}(s, {\bf x}; t, {\bf y})
= \frac{f_{T}^{(\nu,\kappa)}(s, {\bf x}; t,{\bf y})
{\cal N}_{T}^{(\nu, \kappa)}(t, {\bf y})}
{{\cal N}_{T}^{(\nu, \kappa)}(s, {\bf x})}
\label{eqn:gNnk}
\end{equation}
for $0\le s< t \le T, \x, \y \in \W_{N}^{\rm C}$,
where
$f_{T}^{(\nu,\kappa)}(s, {\bf x}; t, {\bf y}) =
\det_{1 \leq i, j \leq N} [
G^{(\nu,\kappa)}_{T}(s,x_i; t,y_j)]$
with (\ref{eqn:meander}) and
${\cal N}_{T}^{(\nu, \kappa)}(t, {\bf x})
= \int_{\W_{N}^{\rm C}} d {\bf y}
f_{T}^{(\nu,\kappa)}(t,{\bf x}; T, {\bf y})$.
Note that
$f_{T}^{(\nu,\kappa)}(s, {\bf x}; t, {\bf y}) =
f^{(\nu)}(t-s, {\bf y}|{\bf x}) 
h^{(\nu,\kappa)}_{T}(t,{\bf y})
/h^{(\nu,\kappa)}_{T}(s,{\bf x})$,
where
$h^{(\nu,\kappa)}_{T}(t,{\bf x}) 
= \prod_{i=1}^N h^{(\nu,\kappa)}_{T}(t,x_i)$.
Since $\lim_{t \to 0} G^{(\nu)}(t, z|w)=\delta(z-w)
{\bf 1}(z \geq 0)$,
$h^{(\nu, \kappa)}_{T}(T, {\bf x})
= \prod_{j=1}^{N} x_{j}^{-\kappa}$ for
$\x \in \W_{N}^{\rm C}$, and then
(\ref{eqn:gNnk}) can be written as
\begin{equation}
g_{T}^{(\nu, \kappa)}(s, {\bf x}; t, {\bf y}) 
= \frac{1}{\widetilde{{\cal N}}^{(\nu, \kappa)}(T-s, {\bf x})}
f^{(\nu)}(t-s, {\bf y}|{\bf x})
\widetilde{{\cal N}}^{(\nu, \kappa)}(T-t, {\bf y})
\label{eqn:gNnk1}
\end{equation}
with
\begin{equation}
\widetilde{{\cal N}}^{(\nu, \kappa)}(t, {\bf x})
= \int_{\W_{N}^{\rm C}} d {\bf y} \
f^{(\nu)}(t, {\bf y}|{\bf x}) 
\prod_{i=1}^{N} y_{i}^{-\kappa}.
\label{eqn:tildeN}
\end{equation}

\begin{lem}
\label{thm:starnk}
Assume that $\nu > -1$ and
$\kappa \in [0,2(\nu+1))$. Let $\x, \y \in \W_{N}^{\rm C}$.
\begin{description}
\item{(i)} \quad For $0 \leq s < t \leq T$,\,
$\displaystyle{
\lim_{T \to \infty} g_{T}^{(\nu, \kappa)}(s,\x; t, \y )
= p^{(\nu)}(s, \x; t, \y)}$.
\item{(ii)} \quad For $0 < t < T$,
\begin{eqnarray}
&& g_{T}^{(\nu, \kappa)}(0, \0; t, {\bf y}) 
\equiv \lim_{|\x| \to 0}
g_{T}^{(\nu, \kappa)}(0, \x; t, {\bf y}) \nonumber\\
&& \quad =\frac{T^{N(N+\kappa-1)/2} t^{-N(N+\nu)}}
{C_{\nu, \kappa}}
 \exp \left\{ -\frac{|{\bf y}|^2}{2t} \right\}
h^{(2\nu+1)}(\y)
\widetilde{{\cal N}}^{(\nu, \kappa)}
(T-t, {\bf y}). \nonumber\\
\label{eqn:gNnk0}
\end{eqnarray}
\item{(iii)} \quad For $T > 0$, \,
${\displaystyle{
\lim_{t \nearrow T} g_{T}^{(\nu, \kappa)}(0, \0; t, \y)
=\frac{T^{-N(N+2\nu+1-\kappa)/2}}{C_{\nu, \kappa}}
 \exp \left\{ -\frac{|{\bf y}|^2}{2T} \right\}
h^{(2\nu+1-\kappa)}(\y)}}$.
\end{description}
\end{lem}
\noindent{\it Proof.} \quad
Using (\ref{eqn:asymfNnu}) for (\ref{eqn:tildeN}), we have
the estimate of $\widetilde{\cal N}^{(\nu,\kappa)}(t,\x)$
in $|\x|/\sqrt{t} \to 0$ as
\begin{eqnarray}
&& \widetilde{\cal N}^{(\nu,\kappa)}(t,\x) 
=\frac{t^{-N(N+2\nu+1)/2}}{C_{\nu}}
\prod_{1 \leq i<j \leq N} \left\{
\left( \frac{x_{j}}{\sqrt{t}} \right)^2
-\left( \frac{x_{i}}{\sqrt{t}} \right)^2 \right\}
\nonumber\\
&& \times
\int_{\W_{N}^{\rm C}}d \y \, 
\prod_{1 \leq k \leq \ell \leq N}
(y_{\ell}^2-y_{k}^2) 
\prod_{m=1}^{N} y_{m}^{2\nu+1-\kappa}
\exp \left\{ - \frac{|\y|^2}{2t} \right\}
\times \left( 1+ {\cal O}\left(
\frac{|\x|}{\sqrt{t}} \right) \right)
\nonumber\\
&=& \frac{t^{-N \kappa/2}C_{\nu, \kappa}}{C_{\nu}}
\prod_{1 \leq i<j \leq N} \left\{
\left( \frac{x_{j}}{\sqrt{t}} \right)^2
-\left( \frac{x_{i}}{\sqrt{t}} \right)^2 \right\}
\times \left( 1+ {\cal O}\left(
\frac{|\x|}{\sqrt{t}} \right) \right),
\label{eqn:asymtildN}
\end{eqnarray}
where we have used a version of Selberg's integral formula
\cite{Sel44,Mac82}
$$
\int_{\R^{N}}d \u \, 
\prod_{1 \leq i < j \leq N}
|u_{j}^2-u_{i}^2|^{2 \gamma}
\prod_{k=1}^{N} |u_{k}|^{2\alpha-1}
e^{-|\u|^2/2}
=2^{\alpha N+\gamma N(N-1)}
\prod_{i=1}^{N} \frac{\Gamma(1+i \gamma)
\Gamma(\alpha+\gamma(i-1))}{\Gamma(1+\gamma)}
$$
by setting $\alpha=\nu+1-\kappa/2$ and
$\gamma=1/2$
(see Equation (17.6.6) in \cite{Meh91}).
By (\ref{eqn:asymfNnu}) and (\ref{eqn:asymtildN}),
(i) and (ii) are obtained.
Since $\lim_{t \to 0} G^{(\nu)}(t, y|x)=\delta(y-x)
{\bf 1}(y \geq 0)$, we have
$\lim_{t \to 0} f^{(\nu)}(t, \y|\x)
=\prod_{i=1}^{N} \delta(y_{i}-x_{i})$
for $\x, \y \in \W_{N}^{\rm C}$.
Then $\lim_{t \to 0} \widetilde{\cal N}^{(\nu,\kappa)}
(t, \x)=\prod_{i=1}^{N} x_{i}^{-\kappa}
{\bf 1}(\x \in \W_{N}^{\rm C})$ and
(iii) is obtained. 
\qed

Now we define the process
$\X^{(\nu, \kappa)}(t)
=(X_{1}^{(\nu, \kappa)}(t), X_{2}^{(\nu, \kappa)}(t), \cdots,
X_{N}^{(\nu,\kappa)}(t)), t \in [0, T],$
as the temporally inhomogeneous diffusion process,
whose transition probability density is given by
(\ref{eqn:gNnk}) for $0 \leq s < t \leq T,
\x, \y \in \W_{N}^{\rm C}$ and
(\ref{eqn:gNnk0}) for $0 < t \leq T,
\y \in \W_{N}^{\rm C}$.
This process solves the stochastic differential equations
$$
dX_{i}^{(\nu, \kappa)}(t)=dB_{i}(t)
+ \left[ \frac{2\nu+1}{2X_{i}^{(\nu, \kappa)}(t)}
+ b_{i}^{(\nu, \kappa)}(T-t, \X^{(\nu, \kappa)}(t)) \right] dt,
\quad t \in [0, T], 1 \leq i \leq N,
$$
where
$b_{i}^{(\nu, \kappa)}(t, \x)= 
(\partial/\partial x_{i})
\ln \widetilde{\cal N}^{(\nu, \kappa)}(t, \x),
\, 1 \leq i \leq N$.

\begin{figure}
\includegraphics[width=.8\linewidth]{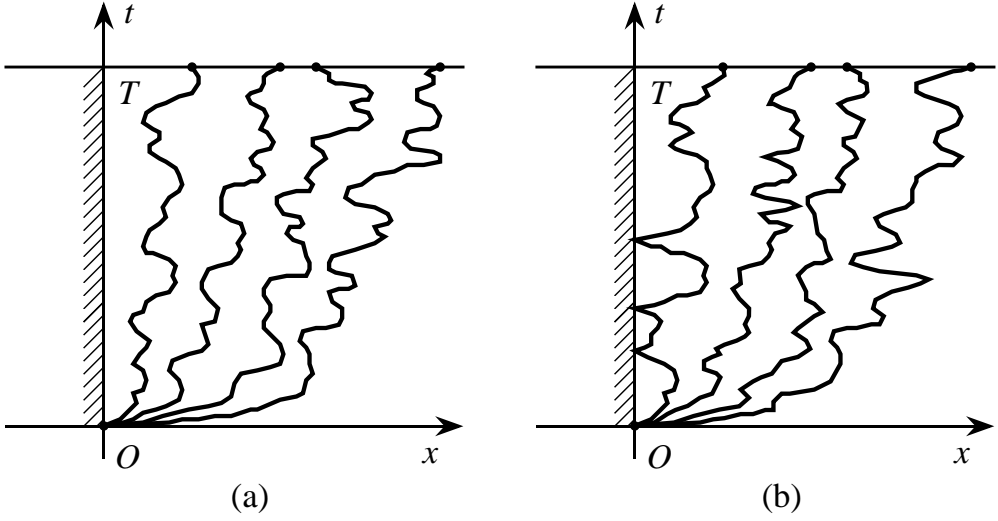}
\caption{(a) Process $\X^{(1/2,1)}(t), t \in [0,T]$
with the initial state $\0$
showing star topology. 
(b) Process $\X^{(-1/2,0)}(t), t \in [0,T]$
with the initial state $\0$
showing star topology. \label{fig:CD}}
\end{figure}

Here we consider the special cases $(\nu, \kappa)=(1/2,1)$
and $(\nu, \kappa)=(-1/2,0)$.
By the definitions (\ref{eqn:NNCD}) and (\ref{eqn:tildeN}),
$\widetilde{\cal N}^{(1/2,1)}(t, \x)=
{\cal N}^{\rm C}(t,x)/\prod_{i=1}^{N} x_{i}$ and
$\widetilde{\cal N}^{(-1/2,0)}(t, \x)=
{\cal N}^{\rm D}(t,x)$,
and then (\ref{eqn:gNnk1}) gives
\begin{eqnarray}
g_{T}^{(1/2,1)}(s,\x; t, \y)&=&
\frac{1}{{\cal N}^{\rm C}(T-s,\x)}
f^{\rm C}(t-s, \y|\x) {\cal N}^{\rm C}(T-t,\y), \nonumber\\
g_{T}^{(-1/2,0)}(s,\x; t, \y)&=&
\frac{1}{{\cal N}^{\rm D}(T-s,\x)}
f^{\rm D}(t-s, \y|\x) {\cal N}^{\rm D}(T-t,\y), \nonumber
\end{eqnarray}
for $0 \leq s < t \leq T, \x, \y \in \W_{N}^{\rm C}$.
That is, we can interpret the process
$\X^{(1/2,1)}(t)$ as the $N$-particle system
of Brownian motions conditioned never to collide with
each other nor with the wall at the origin in
one-dimension during the time-interval $[0,T]$, and
the process $\X^{(-1/2,0)}(t)$ as
the $N$-particle system of reflecting Brownian motions
conditioned never to collide with each other
during the time-interval $[0,T]$, respectively.
The asymptotic forms 
${\cal N}^{\sharp}(t, \x)=
(C[\sharp']/C[\sharp])
h^{\sharp}(\x/\sqrt{t})
\times ( 1+ {\cal O}(|\x|/\sqrt{t}))$
in $|\x|/\sqrt{t} \to 0$ for $\sharp=$C and D
are obtained by (\ref{eqn:asymtildN}), 
and thus we can see the power-laws
of the noncolliding probabilities,
${\cal N}^{\sharp}(t, \x) \sim t^{-\psi[\sharp]}$
as $t \to \infty$
for $\x \in \W_{N}^{\sharp}, \sharp=$ C and D
with the exponents
$\psi[{\rm C}]=N^2/2, 
\psi[{\rm D}]=N(N-1)/2$.
As a corollary of Lemma \ref{thm:starnk}, we have the following.
\begin{cor}
\label{thm:starcor1}
\begin{description}
\item{(i)} \quad
For $0 < t < T$, if $\x \in \W_{N}^{\rm C}$,
\begin{eqnarray}
g_{T}^{(1/2,1)}(0, \0; t, \x) &=&
\frac{T^{\psi[{\rm C}]} t^{-d[{\rm C}]/2}}{C[{\rm C}']}
\exp \left\{ -\frac{|\x|^2}{2t} \right\}
h^{\rm C}(\x) {\cal N}^{\rm C}(T-t, \x), \nonumber\\
g_{T}^{(-1/2,0)}(0, \0; t, \x) &=&
\frac{T^{\psi[{\rm D}]} t^{-d[{\rm D}]/2}}{C[{\rm D}']}
\exp \left\{ -\frac{|\x|^2}{2t} \right\}
h^{\rm D}(\x) {\cal N}^{\rm D}(T-t, \x). \nonumber
\end{eqnarray}
\item{(ii)} \quad
For $T > 0$, if $\x \in \W_{N}^{\rm C}$, 
$$
\lim_{t \nearrow T} g_{T}^{(1/2,1)}(0, \0; t, \x)
= q^{\rm C'}(\x;T), \quad
\lim_{t \nearrow T} g_{T}^{(-1/2,0)}(0, \0; t, \x)
= q^{\rm D'}(\x;T).
$$
\end{description}
\end{cor}
Figure 2 illustrates the processes
$\X^{(1/2,1)}(t)$ and $\X^{(-1/2,0)}(t)$ both starting 
from $\0$. The path-configurations are in star topology.
In the former any particle can not collide with
the wall at the origin, while
in the latter the leftmost particle is
reflected at the wall.
Another corollary of Lemma \ref{thm:starnk} is 
the following.
\begin{cor}
\label{thm:starcor2}
If $\nu \in \N, \x \in \W_{N}^{\rm C}$, \,
$\displaystyle{
\lim_{t \nearrow T} g_{T}^{(\nu, \nu+1)}(0, \0; t, \x)
= q_{\nu}^{\rm chGOE}(\x; T).
}$
\end{cor}
The combination of Lemma \ref{thm:starnk} (i) with
(\ref{eqn:limit1}) and (\ref{eqn:limit2})
of Lemma \ref{thm:xzero},
Corollaries \ref{thm:starcor1} and \ref{thm:starcor2}
implies that $\X^{(1/2,1)}(t)$, $\X^{(-1/2,0)}(t)$
and $\X^{(\nu, \nu+1)}(t)$ with $\nu \in \N$,
all starting from $\0$,
exhibit the transitions from the eigenvalue statistics
of the class C to the class CI,
from the class D to the class associated with
$q^{\rm D'}$ studied in Sec.III.D, and
from chGUE to chGOE, respectively, as time $t$ goes on
from 0 to $T$.
(See Theorem \ref{thm:equiv1} below.)

At the end of this subsection, we discuss the relation between
the temporally homogeneous diffusion process
$\Y^{(\nu)}(t)$ and the temporally inhomogeneous
diffusion process $\X^{(\nu, \kappa)}(t)$
for $t \in [0, T]$.
For a time sequence $t_{0} \equiv 0 < t_{1} < \cdots < 
t_{\ell-1} < t_{\ell} \equiv T < \infty$ with
$\ell \in \{1,2, \cdots\}$, we consider the multi-time
probabilities with the initial state 
$\Y^{(\nu)}(0)=\X^{(\nu, \kappa)}(0)=\x^{(0)}$
$$
P^{\x(0)} \Bigg( \Y^{(\nu)}(t_{1}) \in d \x^{(1)}, \cdots,
\Y^{(\nu)}(t_{\ell}) \in d \x^{(\ell)} \Bigg)
= \prod_{i=1}^{\ell} p^{(\nu)}(t_{i-1}, \x^{(i-1)};
t_{i}, \x^{(i)}) d \x^{(i)},
$$
and
$$
P^{\x(0)} \Bigg( \X^{(\nu, \kappa)}(t_{1}) \in d \x^{(1)}, \cdots,
\X^{(\nu, \kappa)}(t_{\ell}) \in d \x^{(\ell)} \Bigg)
= \prod_{i=1}^{\ell} g_{T}^{(\nu,\kappa)}(t_{i-1}, \x^{(i-1)};
t_{i}, \x^{(i)}) d \x^{(i)},
$$
where we have used the Markov property of the processes.
Assume that $\x^{(0)}=\0$ and 
$\x^{(i)} \in \W_{N}^{\rm C}, 1 \leq i \leq \ell$.
We use the formulae (\ref{eqn:Bessel2}) and (\ref{eqn:gNnk1})
and apply Lemmas \ref{thm:xzero} and \ref{thm:starnk}.
Then we have the equality
$$
\prod_{i=1}^{\ell} g_{T}^{(\nu,\kappa)}(t_{i-1}, \x^{(i-1)};
t_{i}, \x^{(i)})
= T^{N(N+\kappa-1)/2} \frac{C_{\nu}}{C_{\nu,\kappa}}
\prod_{i=1}^{\ell} p^{(\nu)}(t_{i-1}, \x^{(i-1)};
t_{i}, \x^{(i)}) 
\frac{1}{h^{(\kappa)}(\x^{(\ell)})}.
$$
Since this equality holds for arbitrary time sequence
$t_{0}=0 < t_{1} < \cdots < t_{\ell-1}
< t_{\ell}=T < \infty$ with
$\ell \in \{1,2, \cdots\}$, we can conclude the following.

\begin{prop}
\label{thm:Imhof2}
Assume that $\nu > -1, \kappa \in [0, 2(\nu+1))$.
If $\X^{(\nu, \kappa)}(0)=\Y^{(\nu)}(0)=\0$, then
the distribution of the process $\X^{(\nu, \kappa)}(t)$ is
absolutely continuous with that of the
process $\Y^{(\nu)}(t)$ for $t \in [0,T]$
and the Radon-Nikod\'ym density is given by
$$
\frac{P(\X^{(\nu, \kappa)}(\cdot) \in d {\bf w})}
{P(\Y^{(\nu)}(\cdot) \in d {\bf w})}
=  \frac{C_{\nu}T^{N(N+\kappa-1)/2}}
{C_{\nu,\kappa}h^{(\kappa)}({\bf w}(T))}.
$$
\end{prop}

When $N=1$ and $(\nu,\kappa)=(1/2,1)$,
this proposition gives the Imhof relation
between the Brownian meander and the three-dimensional
Bessel process \cite{Imh84}.
The relation stated by (\ref{eqn:Imhof}) \cite{KT02,KT03a} 
and the above proposition are 
regarded as the multivariate generalizations
of the Imhof relation.

\subsection{Brownian bridges and temporally inhomogeneous
matrix-valued processes}

Assume that $\nu \in \N, 0 < T < \infty$. Let 
$B^{\rho}_{ij}(t), \widetilde{B}^{\rho}_{ij}(t),
1 \leq i \leq N+\nu,
1 \leq j \leq N, 0 \leq \rho \leq 3$ be independent one-dimensional
standard Brownian motions. 
For a given matrix $m=(m_{ij}+\sqrt{-1} \widetilde{m}_{ij})
_{1 \leq i \leq N+\nu, 1 \leq j \leq N}$ with
$m_{ij}, \widetilde{m}_{ij} \in \R$, let
$(\beta_{T}^{\rho})_{ij}(t:m_{ij}), 
(\widetilde{\beta}_{T}^{\rho})_{ij}(t:\widetilde{m}_{ij}),
1 \leq i \leq N+\nu,
1 \leq j \leq N, 0 \leq \rho \leq 3$ be the diffusion processes,
which are the solutions of the following stochastic
differential equations:
\begin{eqnarray}
&& (\beta_{T}^{\rho})_{ij}(t:m_{ij}) 
= B_{ij}^{\rho}(t) - \int_0^t 
\frac{(\beta_{T}^{\rho})_{ij}(s:m_{ij})-m_{ij}}{T-s}ds, 
\nonumber\\
&& (\widetilde{\beta}_{T}^{\rho})_{ij}(t:\widetilde{m}_{ij}) 
= \widetilde{B}_{ij}^{\rho}(t) - \int_0^t 
\frac{(\widetilde{\beta}_{T}^{\rho})_{ij}(s:\widetilde{m}_{ij})-
\widetilde{m}_{ij}}
{T-s}ds, \quad
t \in [0,T]. 
\label{eqn:Bbridge}
\end{eqnarray}
The processes
$(\beta_{T}^{\rho})_{ij}(t:m_{ij})$ and
$(\widetilde{\beta}_{T}^{\rho})_{ij}(t:\widetilde{m}_{ij})$
are one-dimensional Brownian bridges of duration $T$ 
both starting from 0
and ending at $m_{ij}$ and $\widetilde{m}_{ij}$,
respectively \cite{Yor92}.
Next for 
$z^{\rho}=(z^{\rho}_{ij})_{1 \leq i, j \leq N} \in {\cal S}(N)$ 
and $\widetilde{z}^{\rho}
=(\widetilde{z}^{\rho}_{ij})_{1 \leq i, j \leq N} \in {\cal A}(N)$,
$0 \leq \rho \leq 3$,
we set
$$
 (s_{T}^{\rho})_{ij}(t: z^{\rho}_{ij})
=
\left\{
   \begin{array}{ll}
      \displaystyle{
      \frac{1}{\sqrt{2}} (\beta_{T}^{\rho})_{ij}(t:\sqrt{2}z^{\rho}_{ij}),
      } & 
   \mbox{if} \ i < j, \\
        & \\
        (\beta_{T}^{\rho})_{ii}(t:z^{\rho}_{ii}), & \mbox{if} \ i=j, \\
   \end{array}\right.  
$$
and
\begin{equation}
(a_{T}^{\rho})_{ij}(t: \widetilde{z}^{\rho}_{ij})
=
\left\{
   \begin{array}{ll}
      \displaystyle{
      \frac{1}{\sqrt{2}} (\widetilde{\beta}_{T}^{\rho})_{ij}(t:
      \sqrt{2} \widetilde{z}^{\rho}_{ij}),
      } & 
   \mbox{if} \ i < j, \\
        & \\
    0, & \mbox{if} \ i=j, \\
    \end{array}\right. 
\label{eqn:saTr}
\end{equation}
with $(s_{T}^{\rho})_{ij}(t: z^{\rho}_{ij})
=(s_{T}^{\rho})_{ji}(t: z^{\rho}_{ji})$ and
$(a_{T}^{\rho})_{ij}(t: \widetilde{z}^{\rho}_{ij})
=-(a_{T}^{\rho})_{ji}(t: \widetilde{z}^{\rho}_{ji})$ for
$i > j$, where
$1 \leq i, j \leq N, 0 \leq \rho \leq 3$ and
$t \in [0,T]$.
We define the matrix-valued processes
$s^{\rho}_{T}(t:z^{\rho})
=((s^{\rho}_{T})_{ij}(t:z^{\rho}_{ij}))_{1 \leq i, j \leq N}
\in {\cal S}(N)$
and
$a^{\rho}_{T}(t:\widetilde{z}^{\rho})
=((a^{\rho}_{T})_{ij}(t:\widetilde{z}^{\rho}_{ij}))_{1 \leq i, j \leq N}
\in {\cal A}(N)$.

In an earlier paper \cite{KT03b}, we considered the $N \times N$
hermitian matrix-valued process
$\Xi_{T}(t)=s^{0}(t)+\sqrt{-1} a_{T}^{0}(t:O), t \in [0,T]$,
where $O$ denotes the $N \times N$ zero matrix and
$s^{0}(t)$ was defined below (\ref{eqn:sar}).
This process is the temporally inhomogeneous 
matrix-valued process realized as an 
interpolation in duration $T$ of the 
first and second processes given in Sec.II.B.
Using the invariance in distribution of the process $\Xi_{T}(t)$
under unitary transformations and our generalized version
of the Imhof relation (\ref{eqn:Imhof}),
we proved the equivalence in distribution of its eigenvalue process
and $\X(t)$ with $\X(0)=\0$.
As a corollary of this equivalence, we derived the
formula for any $\sigma \in \R$,
\begin{equation}
\int_{{\rm U}(N)} d U \,
\exp \left\{ - \frac{1}{2 \sigma^2} {\rm Tr}
(\Lambda_{\x}-U^{\dagger} \Lambda_{\y} U)^2
\right\}
= \frac{C[{\rm A}] \sigma^{d[{\rm A}]}}
{h^{\rm A}(\x) h^{\rm A}(\y)}
\det_{1 \leq i, j \leq N}
\Bigg[ G^{\rm A}(t, y_{j}|x_{i}) \Bigg],
\label{eqn:HCIZ1}
\end{equation}
where $dU$ denotes the Haar measure of ${\rm U}(N)$
normalized as $\int_{\rm U(N)} dU = 1$,
$\Lambda_{\x}={\rm diag}\{x_{1}, \cdots, x_{N}\}$ and
$\Lambda_{\y}={\rm diag}\{y_{1}, \cdots, y_{N}\}$
with $\x, \y \in \W_{N}^{\rm A}$.
This is a stochastic-calculus derivation of
the Harish-Chandra (Itzykson-Zuber) integral formula
\cite{HC57,IZ80}.
In this subsection, we give extensions of this
argument.

As an interpolation of the Laguerre process
(\ref{eqn:Laguerre1}) and the Wishart process
(\ref{eqn:Wishart1}), we define the 
matrix-valued process
$$
\Xi_{T}^{\rm LW}(t)=M_{T}(t)^{\dagger}M_{T}(t),
\quad t \in [0,T],
$$
where
$M_{T}(t)=(B_{ij}^{0}(t)
+\sqrt{-1} (\widetilde{\beta}_{T}^{0})_{ij}(t:O))
_{1 \leq i \leq N+\nu, 1 \leq j \leq N}
\in {\cal M}(N+\nu, N; \C), t \in [0,T]$,
where $O$ denotes the $(N+\nu) \times N$ zero matrix.
Similarly, the interpolations between the processes
(\ref{eqn:XiCD}) and (\ref{eqn:XiCDp1}) are
defined by
\begin{eqnarray}
&& \Xi_{T}^{\rm C}(t)= \sqrt{-1} a_{T}^{0}(t:O) \otimes \sigma_{0}
+ s^{1}(t) \otimes \sigma_{1}
+ s_{T}^{2}(t:O) \otimes \sigma_{2}
+ s^{3}(t) \otimes \sigma_{3}, \nonumber\\
&& \Xi_{T}^{\rm D}(t)= 
\sqrt{-1} a_{T}^{0}(t:O) \otimes \sigma_{0}
+ \sqrt{-1} a_{T}^{1}(t:O) \otimes \sigma_{1}
+ \sqrt{-1} a^{2}(t) \otimes \sigma_{2}
+ s^{3}(t) \otimes \sigma_{3}, \nonumber
\end{eqnarray}
in which $O$ denotes the $N \times N$ zero matrix.
Let $\vkappa^{\rm LW}(t)
=(\kappa^{\rm LW}_{1}(t), \cdots, \kappa^{\rm LW}_{N}(t)),
t \in [0,T]$ be the square roots of the eigenvalues
of $\Xi_{T}^{\rm LW}(t)$ with 
$0 \leq \kappa^{\rm LW}_{1}(t) \leq \cdots
\leq \kappa^{\rm LW}_{N}(t)$ and
$\vlambda^{\sharp}(t)=(\lambda_{1}^{\sharp}(t),
\lambda_{2}^{\sharp}(t), \cdots,
\lambda_{N}^{\sharp}(t))$ be the nonnegative
eigenvalues of $\Xi_{T}^{\rm \sharp}(t)$
with $0 \leq \lambda_{1}^{\sharp}(t)
\leq \cdots \leq \lambda_{N}^{\sharp}(t)$
for $\sharp=$ C and D.
We prove the following equivalence in distribution among the
temporally inhomogeneous diffusion processes.
\begin{thm}
\label{thm:equiv1}
\begin{description}
\item{(i)} \quad
If $\nu \in \N$ and $\X^{(\nu, \nu+1)}(0)=\0$, then
$\vkappa^{\rm LW}(t) =\X^{(\nu, \nu+1)}(t), \,
t \in [0,T]$ in distribution. 
\item{(ii)} \quad
If $\X^{(1/2,1)}(0)=\X^{(-1/2,0)}(0)=\0$, then
$\vlambda^{\rm C}(t)=\X^{(1/2,1)}(t)$ and 
$\vlambda^{\rm D}(t)=\X^{(-1/2,0)}(t), \,
t \in [0,T]$ in distribution.
\end{description}
\end{thm}

\noindent{\it Proof.} \quad
(i) For a given matrix $m=(m_{ij}+\sqrt{-1} \widetilde{m}_{ij})
_{1 \leq i \leq N+\nu, 1 \leq j \leq N}$,
$m_{ij}, \widetilde{m}_{ij} \in \R$, we consider 
${\cal M}(N+\nu, N; \C)$-valued process
$M_{T}(t:m)=((\beta^{0}_{T})_{ij}(t:m_{ij})
+\sqrt{-1} (\widetilde{\beta}^{0}_{T})_{ij}(t:\widetilde{m}_{ij}))
_{1 \leq i \leq N+\nu, 1 \leq j \leq N}, t \in [0,T]$.
From the equations (\ref{eqn:Bbridge}), we have the equation
\begin{equation}
M_{T}(t:m)=M(t)- \int_{0}^{t}
\frac{M_{T}(s:m)-m}{T-s} ds, \quad
t \in [0,T],
\label{eqn:DTeq}
\end{equation}
where $M(t)=(B_{ij}^{0}(t)+\sqrt{-1} \widetilde{B}_{ij}^{0}(t))
_{1 \leq i \leq N+\nu, 1 \leq j \leq N}$.
Let $m_{U}$ and $m_{O}$ be random matrices with distribution
$\mu_{\nu}^{\rm chGUE}(\, \cdot \,;T)$ and
$\mu_{\nu}^{\rm chGOE}(\, \cdot \,;T)$, respectively.
Since $(\beta_{T}^{0})_{ij}(t:\zeta)$ and 
$(\widetilde{\beta}_{T}^{0})_{ij}(t:\zeta),
t \in [0,T]$ are Brownian motions when $\zeta$ is a Gaussian
random variable with variance $T$ independent of
$B_{ij}^{0}(t)$ and $\widetilde{B}_{ij}^{0}(t)$,
if $m_{U}$ and $m_{O}$ are independent of $M(t), t \in [0,T]$,
\begin{equation}
M_{T}(t:m_{U})=M(t), \quad
M_{T}(t:m_{O})=M_{T}(t), \quad t \in [0,T]
\label{eqn:DTeq2}
\end{equation}
in distribution. Moreover,
since the distribution of the process $M(t)$ is invariant
under any transformation $M(t) \to U^{\dagger} M(t) V$,
$U \in {\rm U}(N+\nu), V \in {\rm U}(N)$, the following
lemma is obtained by the equation (\ref{eqn:DTeq}).

\begin{lem}
\label{thm:invariance1}
For any $U \in {\rm U}(N+\nu), V \in {\rm U}(N)$,
$U^{\dagger} M_{T}(t:m) V=M_{T}(t: U^{\dagger} m V), \,
t \in [0,T]$
in distribution
\end{lem}
\noindent By this lemma, if $m$ and $m'$ in
${\cal M}(N+\nu, N; \C)$ have the same radial coordinates,
the processes of radial coordinates
of $M_{T}(t:m)$ and $M_{T}(t:m'), t \in [0,T]$,
are identical in distribution.
Let $\Xi_{T}^{\rm LW}(t:m)=M_{T}^{\dagger}(t:m) M_{T}(t:m)$.
Then the above gives the identification in distribution
of the processes of square roots of eigenvalues
of $\Xi_{T}^{\rm LW}(t:m)$ and $\Xi_{T}^{\rm LW}(t:m'),
t \in [0,T]$.
Now we denote by $P_{T}^{\vkappa}(\cdot)$
the probability distribution of the process of
square roots of eigenvalues of $\Xi^{\rm LW}_{T}(t:m)$
conditioned that the square roots of eigenvalues of $m$ is 
$\vkappa=(\kappa_{1}, \cdots, \kappa_{N})$
with the condition (\ref{eqn:condition2}).
We also denote by $P(\cdot)$ and $P_{T}(\cdot)$
the distributions of the processes of square roots of
eigenvalues of $\Xi(t)=M(t)^{\dagger} M(t)$ and
$\Xi_{T}(t)=M_{T}^{\dagger}(t) M_{T}(t),
t \in [0,T]$, respectively.
The equalities (\ref{eqn:DTeq2}) give 
$$
P(\cdot)=\int_{\W_{N}^{\rm C}} d \vkappa \,
P_{T}^{\vkappa}(\cdot) q_{\nu}^{\rm chGUE}
(\vkappa; T), \quad
P_{T}(\cdot)=\int_{\W_{N}^{\rm C}} d \vkappa \,
P_{T}^{\vkappa}(\cdot) q_{\nu}^{\rm chGOE}
(\vkappa; T).
\label{eqn:relation1}
$$
Then $P_{T}(\cdot)$ and $P(\cdot)$ satisfy the
same relation as the generalized Imhof relation
between $\X^{(\nu, \nu+1)}(t)$ and
$\Y^{(\nu)}(t)$ obtained from 
Proposition \ref{thm:Imhof2} by
setting $\nu \in \N, \kappa=\nu+1$.
Since $P(\cdot)$ is equal to the distribution of the
temporally homogeneous diffusion process $\Y^{(\nu)}(t)$
(see (\ref{eqn:limit1}) of Lemma \ref{thm:xzero}),
we can conclude that
$P_{T}(\cdot)$ is identical to the distribution of the 
process $\X^{(\nu, \nu+1)}(t)$. 

(ii) The second part can be proved by the same
argument as the first part.
For given $y^{\rho}, z^{\rho} \in {\cal S}(N)$, 
$\widetilde{y}^{\rho}$, $\widetilde{z}^{\rho} 
\in {\cal A}(N)$, $0 \leq \rho \leq 3$, put
$Y
= \sqrt{-1} \widetilde{y}^{0} \otimes \sigma_{0}
+ y^{1} \otimes \sigma_{1}
+ y^{2} \otimes \sigma_{2}
+ y^{3} \otimes \sigma_{3}
\in {\cal H}_{2-}(2N)$ and
$Z= 
\sqrt{-1} \widetilde{z}^{0} \otimes \sigma_{0}
+ \sqrt{-1} \widetilde{z}^{1} \otimes \sigma_{1}
+ \sqrt{-1} \widetilde{z}^{2} \otimes \sigma_{2}
+ z^{3} \otimes \sigma_{3} 
\in {\cal H}_{1-}(2N)$.
For these $Y$ and $Z$,
we introduce the temporally
inhomogeneous matrix-valued processes
\begin{eqnarray}
&& \Xi_{T}^{\rm C}(t:Y)
= \sqrt{-1} a_{T}^{0}(t:\widetilde{y}^{0}) \otimes \sigma_{0}
+ s_{T}^{1}(t:y^{1}) \otimes \sigma_{1}
+ s_{T}^{2}(t:y^{2}) \otimes \sigma_{2}
+ s_{T}^{3}(t:y^{3}) \otimes \sigma_{3}, \nonumber\\
&& \Xi_{T}^{\rm D}(t:Z)= 
\sqrt{-1} a_{T}^{0}(t:\widetilde{z}^{0}) \otimes \sigma_{0}
+ \sqrt{-1} a_{T}^{1}(t:\widetilde{z}^{1}) \otimes \sigma_{1}
+ \sqrt{-1} a_{T}^{2}(t:\widetilde{z}^{2}) \otimes \sigma_{2}
+ s_{T}^{3}(t:z^{3}) \otimes \sigma_{3}. \nonumber
\end{eqnarray}
The key lemma \ref{thm:invariance1} of the proof
is replaced by the following.
\begin{lem}
\label{thm:invariance2}
For any $U \in {\rm U}_{2}(2N), 
V \in {\rm U}_{1}(2N)$,
$U^{\dagger} \Xi_{T}^{\rm C}(t: Y) U=
\Xi_{T}^{\rm C}(t: U^{\dagger} Y U)$, and
$V^{\dagger} \Xi_{T}^{\rm D}(t: Z) V=
\Xi_{T}^{\rm D}(t: V^{\dagger} Z V), \,
t \in [0,T]$
in distribution
\end{lem}
For $\sharp=$C and D
we denote by $P_{T}^{\sharp, \vomega}(\cdot)$
the probability distributions of the processes of
nonnegative eigenvalues of $\Xi^{\sharp}_{T}(t:Z)$
conditioned that
the nonnegative eigenvalues of $Z$ is 
$\vomega=(\omega_{1}, \cdots, \omega_{N})$
with (\ref{eqn:condition4}).
We also denote by $P^{\sharp}(\cdot)$ 
and $P^{\sharp}_{T}(\cdot)$
the distributions of the processes of nonnegative
eigenvalues of $\Xi^{\sharp}(t)$ and
$\Xi_{T}^{\sharp}(t),
t \in [0,T]$, respectively. Then we have
the expressions,
$$
P^{\sharp}(\cdot)=\int_{\W_{N}^{\rm C}} d \vomega \,
P_{T}^{\sharp, \vomega}(\cdot) q^{\sharp}(\vomega; T), \quad
P_{T}^{\sharp}(\cdot)=\int_{\W_{N}^{\rm C}} d \vomega \,
P_{T}^{\sharp, \vomega}(\cdot) q^{\sharp'}
(\vomega; T) 
\quad \mbox{for} \quad  \sharp=\mbox{C, D}.
$$
Comparing them with the $(\nu,\kappa)=(1/2,1)$ and
$(\nu, \kappa)=(-1/2,0)$ cases of the generalized
Imhof relations obtained from Proposition \ref{thm:Imhof2},
we have the theorem.
\qed

\vskip 0.5cm

As a corollary of Theorem \ref{thm:equiv1}, the following
integral formulae are derived as proved in Appendix B.

\begin{cor}
\label{thm:HCIZ2}
\begin{description}
\item{(i)} 
Assume $\nu \in \N$ and $\x, \y \in \W_{N}^{\rm C}$.
For any $\sigma \in \R$,
\begin{eqnarray}
&& \int_{{\rm U}(N+\nu) \times {\rm U}(N)} d\mu(U,V) \,
\exp \left\{ - \frac{1}{2\sigma^2}
{\rm Tr} (K_{\x}-U^{\dagger} K_{\y} V)^{\dagger}
(K_{\x}-U^{\dagger} K_{\y} V) \right\} \nonumber\\
&& \qquad =\frac{C_{\nu} \sigma^{N(N+\nu-2)}}
{h^{(\nu)}(\x) h^{(\nu)}(\y)}
\det_{1 \leq i, j \leq N} \left[
e^{-(x_{i}^2+y_{j}^2)/2 \sigma^2}
I_{\nu} \left( \frac{x_{i} y_{j}}{\sigma^2} \right) \right],
\nonumber
\end{eqnarray}
where
$$
  K_{\x} = \left( \matrix{ \widehat{K}_{\x} \cr O} \right), \quad
  K_{\y} = \left( \matrix{ \widehat{K}_{\y} \cr O} \right),
$$ 
with
$\widehat{K}_{\x}={\rm diag}
\{x_{1}, x_{2}, \cdots, x_{N}\},
\widehat{K}_{\y}={\rm diag}
\{y_{1}, y_{2}, \cdots, y_{N}\}$
and $\nu \times N$ zero matrix $O$.
\item{(ii)}
Let $\sharp=\mbox{C, D}$.
For $\x, \y \in \W_{N}^{\rm C}$, $\sigma \in \R$,
$$
 \int_{\widetilde{\rm U}(2N)} dU \, 
\exp  \left\{ -\frac{1}{4 \sigma^2} {\rm Tr}
(\Lambda_{\x}-U^{\dagger} \Lambda_{\y} U)^2 \right\}
\nonumber\\
= \frac{C[\sharp] \sigma^{d[\sharp]}}
{h^{\sharp}(\x) h^{\sharp}(\y)}
\det_{1 \leq i, j \leq N} 
\Big[ G^{\sharp}(\sigma^2, y_{j}|x_{i}) \Big],
$$
where 
$\Lambda_{\x}={\rm diag}\{ x_{1}, x_{2},
\dots, x_{N} \} \otimes \sigma_{3}$,
$\Lambda_{\y}={\rm diag}\{ y_{1}, y_{2},
\dots, y_{N} \} \otimes \sigma_{3},
\widetilde{\rm U}(2N)={\rm U}_{2}(2N)$ for $\sharp={\rm C}$
and $\widetilde{\rm U}(2N)={\rm U}_{1}(2N)$ for $\sharp={\rm D}$.
\end{description}
\end{cor}

They are extensions of the Harish-Chandra (Itzykson-Zuber)
formula (\ref{eqn:HCIZ1}). The formula (i) is found
in \cite{JSV96}.

\subsection{Watermelon topology}

Consider the $N$-particle system of Brownian motions
starting from $\x \in \W_{N}^{\rm A}$ at time $t=0$
and arriving at $\z \in \W_{N}^{\rm A}$ at time $T >0$,
which do not collide with each other during the
time interval $[0,T]$.
We denote by $g^{\rm A, w}(0, \x; t, \y; T,\z)$
the probability density of the state $\y$
at time $t \in [0,T]$.
It is given by
\begin{equation}
g^{\rm A, w}(0, \x; t, \y; T,\z)
= \frac{f^{\rm A}(t, \y|\x) f^{\rm A}(T-t, \z|\y)}
{f^{\rm A}(T, \z|\x)},
\quad \y \in \W_{N}^{\rm A}, t \in [0,T].
\label{eqn:gNTAw1}
\end{equation}
By using (\ref{eqn:asymfA}), we can
obtain the limit
$\displaystyle{g^{\rm A, w}(0,\0; t, \y; T, \0)
=\lim_{|\x| \to 0, |\z| \to 0}
g^{\rm A, w}(0, \x; t, \y; T,\z)}$.
Let $\sigma_{T}(t)= \sqrt{t(1-t/T)}$.
\begin{prop}
\label{thm:watermelon1}
For $\y \in \W_{N}^{\rm A}$, \,
$\displaystyle{
g^{\rm A, w}(0,\0; t, \y; T, \0)
= q^{\rm GUE}(\y; \sigma_{T}(t)^2),
t \in [0,T].
}$
\end{prop}
We denote by $\X^{\rm A, w}(t), t \in [0,T]$,
the temporally inhomogeneous diffusion process,
whose probability density is given by the above.
Its path-configuration on the spatio-temporal plane is
illustrated by Figure 3. Such a pattern
is called {\it watermelon topology} in the polymer
network theory \cite{EG95}.

\begin{figure}
\includegraphics[width=.5\linewidth]{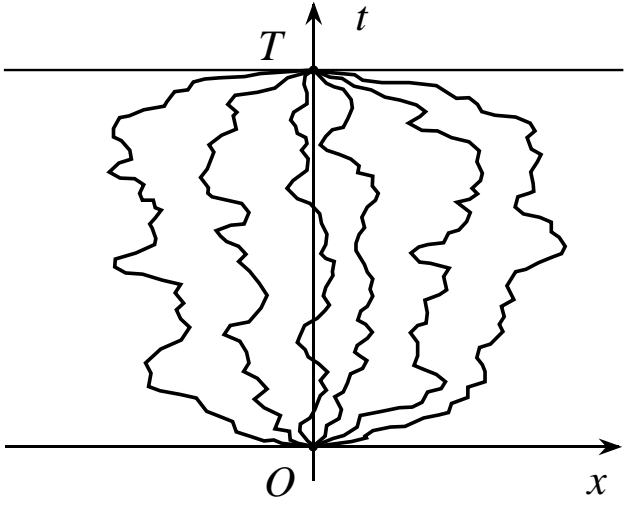}
\caption{Process ${\bf X}^{\rm A, w}(t), t \in [0,T]$, 
showing watermelon topology. \label{fig:watermelon}}
\end{figure}

For $\nu > -1$, similarly to (\ref{eqn:gNTAw1}) we put
$$
g^{(\nu), {\rm w}}(0, \x; t, \y; T, \z)
= \frac{f^{(\nu)}(t, \y|\x) f^{(\nu)}(T-t, \z|\y)}
{f^{(\nu)}(T, \z|\x) }
$$
for $\x, \y, \z \in \W_{N}^{\rm C}, t \in [0,T]$.
By (\ref{eqn:asymfNnu}) we have the following
$\x \to \0$ limit.

\begin{prop}
\label{thm:watermelon2}
For $\nu > -1, \x \in \W_{N}^{\rm C}, t \in [0, T]$,
\begin{eqnarray}
&& g^{(\nu), {\rm w}}(0, \0; t, \x; T, \0)
= \frac{ \sigma_{T}(t)^{-2N(N+\nu)}}
{C_{\nu}} h^{((2\nu+1)/2)}(\x)^{2}
\exp \left\{ -\frac{|\x|^2}{2 \sigma_{T}(t)^2} \right\}.
\nonumber
\end{eqnarray}
In particular, if $\nu \in \N$,
$g^{(\nu),{\rm w}}(0, \0; t, \x; T, \0)=
q_{\nu}^{\rm chGUE}(\x, \sigma_{T}(t)^2)$,
$g^{(1/2),{\rm w}}(0, \0; t, \x; T, \0)=
q^{\rm C}(\x, \sigma_{T}(t)^2)$, and
$g^{(-1/2),{\rm w}}(0, \0; t, \x; T, \0)=
q^{\rm D} (\x, \sigma_{T}(t)^2)$.
\end{prop}
We note that this expression may be formally obtained
by taking $\kappa \to 2(\nu+1)$ limit
of (\ref{eqn:gNnk0}).

\subsection{Banana topology}

For $\varepsilon > 0$, we consider a subspace of $\W_{2N}^{\rm A}$,
$\B_{2N}^{\rm A}(\varepsilon)=\{\x=(x_{1}, x_{2}, \cdots, x_{2N})
\in \W_{2N}^{\rm A}:
x_{2i}=x_{2i-1}+\varepsilon, 1 \leq i \leq N\}$.
For $\x \in \W_{2N}^{\rm A}$, we will use the notation
$\x^{\rm odd}=(x_{1}, x_{3}, \cdots, x_{2N-1})$ and define
${\cal N}^{\rm A, b}(t, \x; \varepsilon)
= \int_{\B_{2N}^{\rm A}(\varepsilon)} d \y^{\rm odd} \,
f^{\rm A}(t, \y|\x)$.
We consider the process, whose transition probability density
is given by
$$
g_{T}^{\rm A, b}(s,\x; t, \y; \varepsilon)
= \frac{f^{\rm A}(t-s, \y|\x) 
{\cal N}^{\rm A, b}(T-t, \y; \varepsilon)}
{{\cal N}^{\rm A, b}(T-s, \x; \varepsilon)},
\quad \x, \y \in \W_{2N}^{\rm A}, 0 \leq s < t \leq T.
$$
This is the $2N$-particle system of noncolliding Brownian
motions in $[0,T]$ conditioned that
the final state at time $t=T$ is in $\B_{2N}^{\rm A}(\varepsilon)$.
Using (\ref{eqn:asymfA}),we have
$$
g_{T}^{\rm A, b}(0, \0; t, \y; \varepsilon)
\equiv \lim_{|\x| \to 0}
g_{T}^{\rm A, b}(0, \x; t, \y; \varepsilon) 
= \left(\frac{t}{T} \right)^{-2N^2}
\frac{ h^{\rm A}(\y) e^{-|\y|^2/2t}
{\cal N}^{\rm A, b}(T-t, \y; \varepsilon)  }
{\int_{\B_{2N}^{\rm A}(\varepsilon)} d \z^{\rm odd} \,
h^{\rm A}(\z) e^{-|\z|^2/2T} }
$$
for $\y \in \W_{2N}^{\rm A}, t \in (0, T]$.
Since $\lim_{t \to 0} f^{\rm A}(0, \y|\x)=\prod_{i=1}^{N}
\delta(x_{i}-y_{i})$,
$\lim_{t \to 0}{\cal N}^{\rm A, b}(t,\x; \varepsilon)
={\bf 1}( \x \in \B_{2N}^{\rm A}(\varepsilon))$, and then
for $\y \in \W_{2N}^{\rm A}$
$$
\lim_{t \nearrow T} g_{T}^{\rm A, b}(0, \0; t, \y; \varepsilon)
= \frac{h^{\rm A}(\y) e^{-|\y|^2/2T}}
{\int_{\B_{2N}^{\rm A}(\varepsilon)} d \z^{\rm odd} \,
h^{\rm A}(\z) e^{-|\z|^2/2T}}
{\bf 1}(\y \in \B_{2N}^{\rm A}(\varepsilon)).
$$
As implied in \cite{MP83} we can take the limit,
$g_{T}^{\rm A, b}(s, \x; t, \y)
= \lim_{\varepsilon \to 0}
g_{T}^{\rm A, b}(s,\x; t, \y; \varepsilon)$,
in the above formulae to have 
\begin{eqnarray}
\label{eqn:gbanana1}
&& g_{T}^{\rm A, b}(s, \x; t, \y)
= \frac{f^{\rm A}(t-s, \y|\x) 
{\cal N}^{\rm A, b}(T-t, \y)}
{{\cal N}^{\rm A, b}(T-s, \x)}, \\
\label{eqn:gbanana2}
&& g_{T}^{\rm A, b}(0, \0; t, \y)
=\left( \frac{T}{2} \right)^{N(2N+1)/2}
\left( \frac{t}{2} \right)^{-2N^2}
\frac{h^{\rm A}(\y)}{C[{\rm A}'']}
e^{-|\y|^2/2t}
{\cal N}^{\rm A, b}(T-t, \y),\\
\label{eqn:gbanana3}
&&
g_{T}^{\rm A, b}(0, \0; T, \y)
= q^{\rm GSE} \left(\y^{\rm odd}; 
\frac{T}{2} \right)
{\bf 1}( \y \in \B_{2N}^{\rm A}),
\end{eqnarray}
for $\x, \y \in \W_{2N}^{\rm A}, 0 \leq s < t <T$,
where
${\cal N}^{\rm A, b}(t, \x)
=\int_{\W_{N}^{\rm A}} d \y \,
f^{\rm A, b}(t, \y|\x)$
with
$$
f^{\rm A, b}(t, \y|\x)=
\det_{1 \leq i \leq 2N, 1 \leq j \leq N}
\left[ G^{\rm A}(t, y_{j}|x_{i}) \quad 
\frac{x_{i}}{t} G^{\rm A}(t, y_{j}|x_{i}) \right]
$$
for $\x \in \W_{2N}^{\rm A}$ and $\y \in \W_{N}^{\rm A}$,
and $\B_{2N}^{\rm A}=\{\x=(x_{1}, x_{2}, \cdots, x_{2N}):
\x^{\rm odd} \in \W_{N}^{\rm A}, 
x_{2i}=x_{2i-1}, 1 \leq i \leq N \}$.
We define the temporally inhomogeneous process
$\X^{\rm A, b}(t), t \in [0,T]$ 
starting from $\0$ or the state in $\W_{2N}^{\rm A}$ and
ending at the state in $\B_{2N}$ as the diffusion
process, whose transition probability density
is given by (\ref{eqn:gbanana1})-(\ref{eqn:gbanana3}).
The path-configuration of $N$ particles
in this version of noncolliding Brownian motions
on the spatio-temporal plane is 
illustrated by Figure 4, which we would like to call
``banana topology".
Important point is that at the final time $t=T$
the particle positions are pairwise degenerated
and distinct positions are identical in distribution
with the Kramers doublets of eigenvalues of
random matrices in GSE as claimed by
(\ref{eqn:gbanana3}).

\begin{figure}
\includegraphics[width=.5\linewidth]{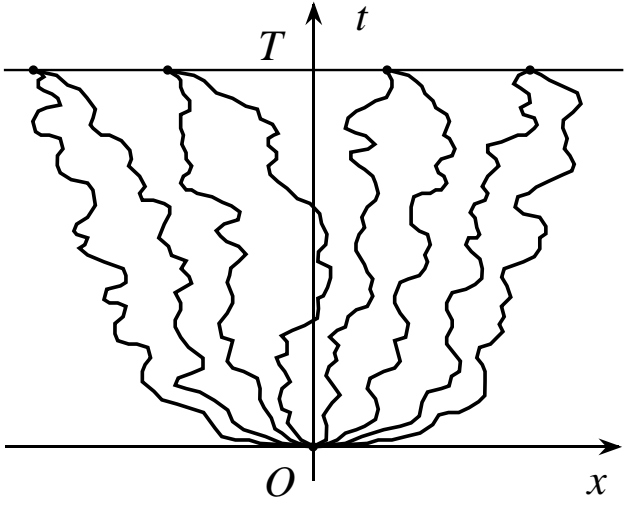}
\caption{Process ${\bf X}^{\rm A, b}(t), t \in [0,T]$,
showing banana topology. \label{fig:banana}}
\end{figure}

Now we consider a $2N \times 2N$ hermitian matrix-valued 
temporally inhomogeneous process defined by
\begin{eqnarray}
\Xi^{\rm b}_{T}(t) &=&
\left\{ s^{0}(t)+\sqrt{-1}a_{T}^{0}(t:O) \right\}
\otimes \sigma_{0} 
+ \left\{ s_{T}^{1}(t:O)+\sqrt{-1}a^{1}(t) \right\}
\otimes \sigma_{1} \nonumber\\
&+& \left\{ s_{T}^{2}(t:O)+\sqrt{-1}a^{2}(t) \right\}
\otimes \sigma_{2}
+  \left\{ s_{T}^{3}(t:O)+\sqrt{-1}a^{3}(t) \right\}
\otimes \sigma_{3},
\label{eqn:XippT}
\end{eqnarray}
where the elements of the $N \times N$ matrices 
$\{s_{T}^{\rho}(t: z^{\rho}),
(a_{T}^{\rho}: \widetilde{z}^{\rho})\}_{\rho=1}^{3}$
are given by (\ref{eqn:saTr}).
By definition, $\Xi^{\rm b}_{T}(T)$ distributes
with the probability density of GSE.
Then the same argument as Theorem \ref{thm:equiv1}
may prove the following.

\begin{thm}
\label{thm:equiv2}
Let $\vlambda(t)=(\lambda_{1}(t), \lambda_{2}(t), \cdots,
\lambda_{2N}(t))$ be the eigenvalues of the process (\ref{eqn:XippT})
with $\lambda_{1}(t) \leq \lambda_{2}(t) \leq \cdots
\leq \lambda_{2N}(t)$.
If $\X^{\rm A, b}(0)=\0$, then
$\vlambda(t)=\X^{\rm A, b}(t), \, t \in [0,T]$
in distribution
\end{thm}

As a corollary of this theorem, we will have the following
version of Harish-Chandra formula, which is found as
Equation (3.46) in \cite{MP83}.

\begin{cor}
\label{eqn:HCIZ3}
Let $\x=(x_{1}, x_{2}, \cdots, x_{2N}) \in \W_{2N}^{\rm A}$,
$\y=(y_{1}, y_{2}, \cdots, y_{N}) \in \W_{N}^{\rm A}$.
For any $\sigma \in \R$
$$
\int_{U(2N)} dU \, 
\exp \left\{ - \frac{1}{2 \sigma^2}
{\rm Tr}(\Lambda_{\x}-U^{\dagger} \Lambda_{y} U)^2 \right\}
\nonumber\\
= \frac{C_{2N}[{\rm A}] \sigma^{(2N)^2}}
{h^{\rm A}(\x) h^{\rm A}(\y)^4}
f^{\rm A, b}(\sigma^2, \y|\x),
$$
where $\Lambda_{\x}={\rm diag}\{x_{1}, x_{2},
\cdots, x_{2N}\}$,
$\Lambda_{\y}={\rm diag}
\{y_{1}, y_{2}, \cdots, y_{N}\} \otimes \sigma_{0}$,
and
$C_{2N}[{\rm A}]=(2\pi)^{N} \prod_{i=1}^{2N} \Gamma(i)$.
\end{cor}

It is easy to see by the same argument that
the transition probability density given below
defines the temporally inhomogeneous
diffusion process $\X^{(\nu, \kappa), {\rm b}}(t),
t \in [0,T], \nu >-1, \kappa \in [0, 2(\nu+1))$,
associated with $\X^{(\nu, \kappa)}$,
which shows the banana topology:
Let
$$
f^{(\nu), {\rm b}}(t, \y|\x)
=\det_{1 \leq i \leq 2N, 1 \leq j \leq N}
\Bigg[ G^{(\nu)}(t, y_{j}|x_{i}) \quad
 G_{y}^{(\nu)}(t, y_{j}|x_{i}) \Bigg]
$$
for $\x \in \W_{2N}^{\rm C}, \y \in \W_{N}^{\rm C}$, 
where
$G_{y}^{(\nu)}(t, y|x)=(\partial/\partial y) G^{(\nu)}(t,y|x)$,
and let
$\widetilde{\cal N}^{(\nu, \kappa), {\rm b}}(t, \x)$
$= \int_{\W_{N}^{\rm C}} d\y \,
f^{(\nu), {\rm b}}(t, \y|\x)$
$\prod_{i=1}^{N} y_{i}^{-\kappa}$ 
for $\x \in \W_{2N}^{\rm C}$.
Then
\begin{eqnarray}
&&g_{T}^{(\nu, \kappa), {\rm b}}(s, \x; t, \y)
= \frac{f^{(\nu)}(t-s, \y|\x)
\widetilde{\cal N}^{(\nu, \kappa), {\rm b}}(T-t, \y)}
{\widetilde{\cal N}^{(\nu, \kappa), {\rm b}}(T-s, \x)},
\nonumber\\
&& g_{T}^{(\nu, \kappa), {\rm b}}(0, \0; t, \y)
= \frac{ 2^{N(4N+4\nu-1)} T^{2N^2} t^{-2N(2N+\nu)}}
{\hat{C}_{\nu}} h^{(2\nu+1)}(\y)
e^{-|\y|^2/2t}
\widetilde{\cal N}^{(\nu, \kappa), {\rm b}}(T-t, \y), 
\nonumber\\
\label{eqn:gnkb3}
&&g_{T}^{(\nu, \kappa), {\rm b}}(0, \0; T, \y)
= \frac{1}{\hat{C}_{\nu}}
\left( \frac{2}{T} \right)^{2N(N+\nu)}
h^{((4 \nu-2\kappa+3)/4)}(\y^{\rm odd})^{4}
e^{-|\y^{\rm odd}|^2/T}
{\bf 1}(\y \in \B_{2N}^{\rm C}), 
\end{eqnarray}
for $\x, \y \in \W_{2N}^{\rm C}, 0 \leq s < t < T$,
where
$\hat{C}_{\nu}=2^{N(2N+2\nu-1)} 
\prod_{i=1}^{N} \left\{\Gamma(2i) \Gamma(2(i+\nu)) \right\}$ and
$\B_{2N}^{\rm C}=\{\x=(x_{1}, x_{2}, \cdots, x_{2N}):
\x^{\rm odd} \in \W_{N}^{\rm C}, 
x_{2i}=x_{2i-1}, 1 \leq i \leq N \}$.
We should notice that (\ref{eqn:gnkb3})
includes the following special cases.
\begin{eqnarray}
&& g_{T}^{(\nu, 0), {\rm b}}(0, \0; T, \y)
= q_{\nu}^{\rm chGSE}\left( \y^{\rm odd};
\frac{T}{2} \right) 
{\bf 1}(\y \in \B_{2N}^{\rm C}) \quad
\mbox{for} \quad \nu \in \N, \nonumber\\
&& g_{T}^{(-1/2, 0), {\rm b}}(0, \0; T, \y)
= q^{\rm DIII}\left( \y^{\rm odd};
\frac{T}{2} \right) 
{\bf 1}(\y \in \B_{2N}^{\rm C}). \nonumber
\end{eqnarray}
Here
$$
q_{\nu}^{\rm chGSE}(\vkappa; t)
=\frac{t^{-2N(N+\nu)}}{\hat{C}_{\nu}}
\exp \left\{-\frac{|\vkappa|^2}{2t} \right\}
\prod_{1 \leq i < j \leq N}
(\kappa_{j}^2-\kappa_{i}^2)^{4} 
\prod_{k=1}^{N} \kappa_{k}^{4 \nu+3}
$$
is the probability density of
the $N$ distinct square roots 
$\vkappa=(\kappa_{1}, \kappa_{2}, \cdots,
\kappa_{N})$ with (\ref{eqn:condition2})
of the eigenvalues of $M^{\dagger} M$
conditioned that $M$ is a $2N \times 2N$ random matrices in the
chiral Gaussian symplectic ensemble
(chGSE) with variance $t$ \cite{VZ93,Ver94,JSV96,SV98}, and
$$
q^{\rm DIII}(\vomega; t)
= \frac{t^{-d[{\rm D}'']/2}}{C[{\rm D}'']} 
\exp \left\{ - \frac{|\vomega|^2}{2t} \right\}
\prod_{1 \leq i < j \leq N}(\omega_{j}^2-\omega_{i}^2)^{4}
\prod_{k=1}^{N} \omega_{k}
$$
with $d[{\rm D}'']=2N(2N-1),
C[{\rm D}'']=\hat{C}_{-1/2}=2^{2N(N-1)} \prod_{i=1}^{N}
\Gamma(2i) \Gamma(2i-1)$
is the probability density of
the nonnegative and distinct eigenvalues 
$\vomega=(\omega_{1}, \omega_{2}, \cdots,
\omega_{N})$ with (\ref{eqn:condition4})
of $4N \times 4N$ matrices in the
ensemble in the symmetry class DIII studied 
by Altland and Zirnbauer \cite{Zir96,AZ96,AZ97}.
(Strictly speaking, it is the DIII-even case.
The DIII-odd case is obtained by setting 
$\nu=1/2, \kappa=0$ in (\ref{eqn:gnkb3}).)
The above implies that $\X^{(\nu, 0), {\rm b}}(t)$ with $\nu \in \N$
and $\X^{(-1/2,0), {\rm b}}(t)$,
both starting from $\0$,
exhibit the transitions from the eigenvalue statistics
of chGUE to chGSE and 
from the class D to the class DIII,
respectively, as time $t$ goes on from 0 to $T$.

A lengthy but explicit expression for the $4N \times 4N$
hermitian matrix-valued process corresponding to
$\X^{(-1/2,0), {\rm b}}(t)$ is given as
\begin{eqnarray}
\Xi_{T}^{\rm D, b}(t) &=& \sum_{\rho=0}^{2} \Bigg\{
\sqrt{-1} a_{T}^{0 \rho}(t:O) \otimes (\sigma_{0} \otimes \sigma_{\rho})
+ \sqrt{-1} a^{1 \rho}(t) \otimes (\sigma_{1} \otimes \sigma_{\rho})
\nonumber\\
&& \qquad
+s_{T}^{2 \rho}(t:O) \otimes (\sigma_{2} \otimes \sigma_{\rho})
+\sqrt{-1} a^{3 \rho}(t) \otimes (\sigma_{3} \otimes \sigma_{\rho}) \Bigg\}
\nonumber\\
&&\quad +\Bigg\{ s^{0 3}(t) \otimes (\sigma_{0} \otimes \sigma_{3})
+s^{1 3}(t) \otimes (\sigma_{1} \otimes \sigma_{3})  \nonumber\\
&& \qquad 
+ \sqrt{-1} a_{T}^{23}(t:O) \otimes (\sigma_{2} \otimes \sigma_{3})
+ s_{T}^{33}(t:O) \otimes (\sigma_{3} \otimes \sigma_{3}) \Bigg\},
\nonumber
\end{eqnarray}
$t \in [0,T]$,
where $s^{\mu \rho}(t), s^{\mu \rho}_{T}(t:O) \in {\cal S}(N)$
and $a^{\mu \rho}(t), a^{\mu \rho}_{T}(t:O) \in {\cal A}(N)$,
$t \in [0,T]$, 
are defined similarly to (\ref{eqn:sar}) and (\ref{eqn:saTr}).
Identification of its eigenvalue process with
$\X^{(-1/2,0), {\rm b}}(t)$ gives the following version
of Harish-Chandra integral,
\begin{eqnarray}
&& \int_{U_{1}(4N)} dU \, \exp \left\{
- \frac{1}{4 \sigma^2} {\rm Tr} (\Lambda_{\x}-U^{\dagger} \Lambda_{\y} U)^2
\right\} \nonumber\\
&=& \frac{C_{2N}[{\rm D}] \sigma^{2N(4N+1)}}
{h^{\rm D}(\x) h^{(1/4)}(\y)^4}
\det_{1 \leq i \leq 2N, 1 \leq j \leq N}
\left[ G^{\rm D}(\sigma^2, y_{j}|x_{i}) \quad
\frac{x_{i}}{\sigma^2} G^{\rm C}(\sigma^2, y_{j}|x_{i}) \right]
\nonumber
\end{eqnarray}
for any $\sigma \in \R$,
$\x=(x_{1}, x_{2}, \cdots, x_{2N}) \in \W_{2N}^{\rm C},
\y=(y_{1}, y_{2}, \cdots, y_{N}) \in \W_{N}^{\rm C}$, where
$\Lambda_{\x}={\rm diag}\{x_{1}, x_{2}, \cdots, x_{2N}\}
\otimes \sigma_{3}$,
$\Lambda_{\y}={\rm diag}\{y_{1}, y_{2}, \cdots, y_{N}\}
\otimes (\sigma_{3} \otimes \sigma_{0})$,
and 
$C_{2N}[{\rm D}]=(\pi/2)^{N}$ $\prod_{i=1}^{2N} \Gamma(2i-1)$.

\section{\large CONCLUDING REMARKS}

In the present paper we showed that the eigenvalue processes
of GUE, chGUE, the class C, and the class D are 
realized by the temporally homogeneous noncolliding diffusion
processes and then the temporally inhomogeneous noncolliding 
diffusion processes were introduced, which
exhibit the transitions in distribution
from the eigenvalue statistics of GUE to GOE,
GUE to GSE, chGUE to chGOE, chGUE to chGSE,
the class C to the class CI, and
the class D to the class DIII.
They are obtained as the special cases of the noncolliding
systems of the Brownian motions and those of Yor's generalized
meanders.
These inhomogeneous processes are identified with the eigenvalue
processes of the inhomogeneous matrix-valued processes, some of
which are regarded as the stochastic versions of two-matrix models
studied by Pandey and Mehta \cite{PM83,MP83} as demonstrated
in \cite{KT02,KTNK03}.
We would like to put emphasis on the fact that
in order to prove the identification
we have not used any results by Pandey and Mehta,
but used the generalized versions of Imhof relations
((\ref{eqn:Imhof}) and Proposition \ref{thm:Imhof2}).
Therefore we can give the proof for the Harish-Chandra
(Itzykson-Zuber)-type integration formulae as corollaries.
The present study suggests several open problems.
Here we list up some of them.
\begin{description}
\item{(i)} \quad It does not seem to be possible to realize
the eigenvalue processes of the random matrix ensembles
different from GUE, chGUE, the class C and the class D
by any temporally homogeneous noncolliding systems of
diffusion particles. Is it possible to realize them as
the temporally homogeneous diffusion processes
with some conditions additional to the simple
noncolliding condition ?
\item{(ii)} \quad Norris, Rogers and Williams \cite{NRW86}
studied other matrix-valued process called
Dynkin's Brownian motion
$\widetilde{\Xi}(t)=G(t)^{T} G(t)$ with
$\partial G(t)=(\partial B(t))G(t)$, where
$\partial$ denotes the Stratonovich differential;
$x \partial y=x dy +dx dy/2$ for
continuous semimartingales $x, y$.
They showed that the eigenvalues
of $\widetilde{\Xi}(t)$ are also noncolliding systems
and derived the stochastic differential equations
similar to (\ref{eqn:Dyson1})
for the logarithms of the eigenvalues.
As mentioned by Bru (see Remark 2 in \cite{Bru91}),
$G(t)$ is a matrix-version of 
{\it multiplicative Brownian motion}
in a sense, while $B(t)$ is the ordinary additional Brownian
motion. Can we discuss (the logarithms of )
the eigenvalue processes using the random matrix theory
and noncolliding diffusion processes as well ?
\item{(iii)} \quad
In the non-hermitian random matrix ensembles, eigenvalues
are distributed on the complex plane
\cite{Gin65,Ede97}.
Is it meaningful to consider the stochastic version of
non-hermitian random matrix theory in the sense
of Dyson \cite{Dys62a} ?
\end{description}

For the temporally inhomogeneous noncolliding Brownian
motions $\X(t), t \in [0,T]$ with $\X(0)=\0$,
the determinantal expressions for the multi-time
correlation functions were determined by
Nagao and the present authors using the self-dual
quaternion matrices \cite{NF99,FNH99,Nag01}
and the scaling limits of the infinite particles
$N \to \infty$ and the infinite time-interval
$T \to \infty$ were investigated \cite{NKT03,KNT03}.
Recently Nagao reported the similar calculation
on the process, which corresponds to the process
$\X^{(1/2,1)}$ in the present paper \cite{Nag03}.
Calculation of the multi-time correlations
for the general process $\X^{(\nu, \kappa)}(t)$
is now in progress and the study of the infinite
particle systems will be reported elsewhere
\cite{KT04}.


\section*{\large ACKNOWLEDGEMENTS}
One of the authors (M.K.) thanks Taro Nagao and
Takahiro Fukui for useful discussion on random matrix
theory and representation theory.
He also thanks the Yukawa Institute for Theoretical 
Physics at Kyoto University, where some application of
the present work was discussed during the 
workshop YITP-W-03-18 on
``Stochastic models in statistical mechanics."
\vskip 1cm

\clearpage

\noindent
{\Large {\bf APPENDICES}}
\vskip 0.5cm

\appendix
\renewcommand{\theequation}{\thesection.\arabic{equation}}
\setcounter{equation}{0}
\section{\large SCHUR FUNCTION EXPANSIONS OF DETERMINANTS}

Any sequence $\mu=(\mu_{1}, \mu_{2}, \cdots, \mu_{N}, \cdots)$
of nonnegative integers in decreasing order
$\mu_{1} \geq \mu_{2} \geq \cdots \geq \mu_{N} \geq \cdots$ is
called a partition. The non-zero $\mu_{i}$ in $\mu$
are called the parts of $\mu$ and the number of parts
is the length of $\mu$ denoted by $\ell(\mu)$.
For each partition $\mu$ with $\ell(\mu) \leq N$,
the Schur function defined by
$s_{\mu}(\x)= \det_{1 \leq i, j \leq N}
(x_{i}^{\mu_{j}+N-j})/
\det_{1 \leq i, j \leq N}(x_{i}^{N-j})$
gives a symmetric polynomial of order 
$|\mu|=\sum_{i=1}^{N} \mu_{i}$ in $N$ variables
$x_{1}, x_{2}, \cdots, x_{N} \in \C$.
Note that the denominator is the Vandermonde determinant and
$\det_{1 \leq i, j \leq N}(x_{i}^{N-j})=
(-1)^{N(N-1)/2} h^{\rm A}(\x)$
\cite{Mac95,Ful97,Sta99}.
We can prove the following expansion formulae of the determinants
with the bases of the Schur functions \cite{Bal00,Bal02,KO01}.
\begin{eqnarray}
\frac{\displaystyle{\det_{1 \leq i, j \leq N} 
 \Big[ e^{x_{i} y_{j}} \Big] }}
 {h^{\rm A}(\x) h^{\rm A}(\y)}
 &=& \sum_{\mu:\ell(\mu) \leq N} a_{\mu} s_{\mu}(\x) s_{\mu}(\y),
\nonumber\\
\frac{\displaystyle{
\det_{1 \leq i, j \leq N} 
\Big[ I_{\nu}(2 \sqrt{x_{i} y_{j}}) \Big] }}
{\displaystyle{ 
\left\{ \prod_{i=1}^{N} x_{i}^{\nu/2} y_{i}^{\nu/2} \right\}
 h^{\rm A}(\x) h^{\rm A}(\y)}}
&=& \sum_{\mu: \ell(\mu) \leq N} b^{(\nu)}_{\mu}
s_{\mu}(\x) s_{\mu}(\y),
\nonumber
\end{eqnarray}
where
$a_{\mu}=1/\prod_{i=1}^{N}
\Gamma(\mu_{i}+N-i+1)$ and 
$b^{(\nu)}_{\mu}=1/\{\prod_{i=1}^{N}
\Gamma(\mu_{i}+N-i+1) \Gamma(\nu+\mu_{i}+N-i+1)\}$. 
Since $s_{\mu}(\0)={\bf 1}(\mu=\0)$ with
$\0=(0,0, \cdots, 0) \in \N^{N}$,
from the above formulae, we have the following
asymptotics of the determinants.
As $|\x| \to 0$,
\begin{eqnarray}
\label{eqn:Schur1}
&& \det_{1 \leq i, j \leq N} \Big[ e^{x_{i} y_{j}} \Big]
= \frac{h^{\rm A}(\x) h^{\rm A}(\y)}{\prod_{i=1}^{N} \Gamma(i)}
\times ( 1+{\cal O}\left( |\x| \right) ), \\
\label{eqn:Schur2}
&& \det_{1 \leq i, j \leq N} 
\Big[ I_{\nu}(2 \sqrt{x_{i} y_{j}}) \Big]
= \left\{ \prod_{i=1}^{N} x_{i}^{\nu/2} y_{i}^{\nu/2} \right\}
\frac{h^{\rm A}(\x) h^{\rm A}(\y)}
{\prod_{j=1}^{N}  \Gamma(j) \Gamma(\nu+j) }
\times (1+{\cal O}(|\x|) ).
\end{eqnarray}

\setcounter{equation}{0}
\section{\large PROOF OF COROLLARY \ref{thm:HCIZ2}}

By (\ref{eqn:gNnk0}) of Lemma \ref{thm:starnk} (ii),
\begin{eqnarray}
&& g_{T}^{(\nu,\kappa)}(0, \0;t, \y)=
\frac{T^{N(N+\kappa-1)/2}t^{-N(N+\nu)}}
{C_{\nu, \kappa}}
\exp \left\{ - \frac{|\y|^2}{2t} \right\}
h^{(2\nu+1)}(\y) \nonumber\\
&& \times \int_{\W_{N}^{\rm C}} d \z \,
\det_{1 \leq i, j \leq N} \left[
\frac{z_{j}^{\nu+1}}{y_{i}^{\nu}}\frac{1}{T-t}
\exp \left\{ -\frac{y_{i}^2+z_{j}^2}{2(T-t)} \right\}
I_{\nu} \left( \frac{y_{i} z_{j}}{T-t} \right) \right]
\times \prod_{k=1}^{N} z_{k}^{-\kappa} \nonumber\\
&=& \frac{T^{N(N+\kappa-1)/2} t^{-N(N+\nu)}}
{(T-t)^{N} C_{\nu, \kappa}}
\left( \frac{T}{t} \right)^{N(\nu+1-\kappa)}
h^{(\nu+1)}(\y) \int_{\W_{N}^{\rm C}} d \z \,
\exp \left\{ - \frac{T}{2t^2} \left(\frac{t}{T}\right)^2
|\z|^2 \right\} \nonumber\\
&& \qquad \times
\det_{1 \leq i, j \leq N} \left[ \exp \left\{
-\frac{T}{2t(T-t)} \left( y_{i}^2+\frac{t^2}{T^2}z_{j}^2 \right)
\right\}
I_{\nu} \left( \frac{T}{t(T-t)} y_{i} \times \frac{t}{T} z_{j} \right)
\right] \prod_{\ell=1}^{N} \left( \frac{t}{T} z_{\ell} \right)
^{\nu+1-\kappa}. \nonumber
\end{eqnarray}
Setting $(t/T)z_{i}=a_{i}, 1 \leq i \leq N$,
$t(1-t/T)=\sigma^2$ and $T/t^2=\alpha$, we have
\begin{eqnarray}
&& g_{T}^{(\nu,\kappa)}(0,\0; t, \y)=
\frac{\sigma^{-2N} \alpha^{N(N+2\nu-\kappa+1)/2}}
{C_{\nu, \kappa}} h^{(\nu+1)}(\y)
\nonumber\\
&& \qquad \times
\int_{\W_{N}^{\rm C}} d \a \,
e^{-\alpha|\a|^2/2}
\det_{1 \leq i, j \leq N} \left[ 
e^{-(y_{i}^2+a_{j}^2)/2 \sigma^2} 
I_{\nu} \left( \frac{y_{i} a_{j}}{\sigma^2} \right) \right]
\prod_{\ell=1}^{N} a_{\ell}^{\nu+1-\kappa}.
\label{eqn:hc1}
\end{eqnarray}
\noindent{\it Proof of (i).} 
We write the transition probability density of the process
$M_{T}(t)$ by $Q_{T}(s, m_{1}; t, m_{2})$, $0 \leq s < t \leq T$,
for $m_{1}, m_{2} \in {\cal M}(N+\nu,N; \C)$.
Then by Theorem \ref{thm:equiv1} (i) 
and the fact (\ref{eqn:chGUEJ}),
\begin{equation}
g_{T}^{(\nu,\nu+1)}(0, \0; t, \y)=
\frac{(2\pi)^{N(N+\nu)}}{C_{\nu}} 
h^{((2\nu+1)/2)}(\y)^2
\int_{{\rm U}(N+\nu) \times {\rm U}(N)}
d \mu(U,V) \, Q_{T}(0, O; t, U^{\dagger} K_{\y} V).
\label{eqn:hc2}
\end{equation}
We introduce the ${\cal M}(N+\nu, N; \C)$-valued process
$M^{(1)}(t)=(m_{ij}^{(1)}(t))_{1 \leq i \leq N+\nu,
1 \leq j \leq N}$ and 
the ${\cal M}(N+\nu, N; \R)$-valued process
$M^{(2)}(t)=(m_{ij}^{(2)}(t))_{1 \leq i \leq N+\nu,
1 \leq j \leq N}$, whose elements are defined by
$$
m_{ij}^{(1)}(t)=B_{ij}^{0}(t)-\frac{t}{T} B_{ij}^{0}(T)
+\sqrt{-1} (\widetilde{\beta}_{T}^{0})_{ij}(t) \quad
\mbox{and} \quad
m_{ij}^{(2)}(t)=\frac{t}{T} B_{ij}^{0}(T).
$$
Then $M_{T}(t)=M^{(1)}(t)+M^{(2)}(t)$.
Note that $\{B_{ij}^{0}(t)-(t/T)B_{ij}^{0}(T)\}$ are Brownian
bridges of duration $T$ starting at 0 and ending at 0, which
are independent of $(t/T)B_{ij}^{0}(T)$.
Hence $M^{(1)}(t)$ is in the chiral GUE distribution
and $M^{(2)}(t)$ in the chiral GOE distribution,
where $M^{(1)}(t)$ and $M^{(2)}(t)$ are
independent from each other.
Since $E[m^{(1)}_{ii}(t)^2]=\sigma^2$
and $E[m^{(2)}_{ii}(t)^2]=1/\alpha$,
$Q_{T}(0,O; t, M)$ for $M \in {\cal M}(N+\nu,N; \C)$
can be written as
\begin{eqnarray}
&& Q_{T}(0,O;t,M)=
\int_{{\cal M}(N+\nu, N; \R)} {\cal V}(dB)
\mu_{\nu}^{\rm chGOE}(B; 1/\alpha)
\mu_{\nu}^{\rm chGUE}(M-B; \sigma^2) \nonumber\\
&=& \frac{\alpha^{N(N+\nu)/2} \sigma^{-N(N+\nu)}}
{C_{\nu,\nu+1}(2\pi)^{N(N+\nu)}}
\int_{\W_{N}^{\rm C}} d \a \,
h^{(\nu)}(\a) 
e^{-\alpha|\a|^2/2 -
{\rm Tr} (M-K_{\a})^{\dagger} (M-K_{\a})/2 \sigma^2},
\label{eqn:hc3}
\end{eqnarray}
where we have used the fact (\ref{eqn:chGOEJ}) and
the formulae (\ref{eqn:Laguerre4}), (\ref{eqn:Wishart4}).
Combining (\ref{eqn:hc1}) with $\kappa=\nu+1$,
(\ref{eqn:hc2}) and (\ref{eqn:hc3}), we have
\begin{eqnarray}
&& \frac{C_{\nu} \sigma^{N(N+\nu-2)}}{h^{(\nu)}(\y)}
\int_{\W_{N}^{\rm C}} d \a \, 
e^{-\alpha |\a|^2/2 }
\det_{1 \leq i, j \leq N} \left[
\exp \left\{ -\frac{y_{i}^2+a_{j}^2}{2 \sigma^2} \right\}
I_{\nu} \left( \frac{y_{i} a_{j}}{\sigma^2} \right) \right]
\nonumber\\
&=& \int_{\W_{N}^{\rm C}} d \a \, h^{(\nu)}(\a)
e^{-\alpha |\a|^2/2}
\int_{{\rm U}(N+\nu) \times {\rm U}(N)}
d\mu(U,V) \, 
e^{- {\rm Tr}(U^{\dagger} K_{\y} V-K_{\a})^{\dagger}
(U^{\dagger} K_{\y}V-K_{\a})/2 \sigma^2}.
\nonumber
\end{eqnarray}
Since, for each $\sigma \in \R$, this equality holds
for any $\alpha >0$, we have the formula (i). \\
\noindent{\it Proof of (ii).} 
By setting $(\nu,\kappa)=(1/2,1)$ and $(\nu,\kappa)=(-1/2,0)$
in (\ref{eqn:hc1}) we have the expressions
for $\x, \y \in \W_{N}^{\rm C}$,
\begin{eqnarray}
g^{(1/2,1)}_{T}(0,\0;t,\x)
= \frac{\alpha^{N(N+1)/2}}{C[{\rm C}']}
h^{\rm C}(\x) \int_{\W_{N}^{\rm C}} d \a \,
e^{-\alpha|\a|^2/2}
\det_{1 \leq i, j \leq N} \Bigg[
G^{\rm C}(\sigma^2, a_{j}|x_{i}) \Bigg], \nonumber\\
g^{(-1/2,0)}_{T}(0,\0;t,\y)
= \frac{\alpha^{N^2/2}}{C[{\rm D}']}
h^{\rm D}(\y) \int_{\W_{N}^{\rm D}} d \a \,
e^{-\alpha|\a|^2/2}
\det_{1 \leq i, j \leq N} \Bigg[
G^{\rm D}(\sigma^2, a_{j}|y_{i}) \Bigg]. \nonumber
\end{eqnarray}
Following the same argument with the proof of (i) and
using the equalities (\ref{eqn:CDJ}) and (\ref{eqn:CDpJ}),
the formulae (ii) are proved.
\qed

\clearpage
\footnotesize


\begin{thebibliography}{99}
\bibitem{AZ96}
A. Altland and M. R. Zirnbauer,
``Random matrix theory of a chaotic Andreev quantum dot,"
Phys. Rev. Lett. {\bf 76}, 3420-3423 (1996).

\bibitem{AZ97}
A. Altland and M. R. Zirnbauer, 
``Nonstandard symmetry classes
in mesoscopic normal-superconducting hybrid structure,"
Phys. Rev. B {\bf 55}, 1142-1161 (1997).

\bibitem{Bal00}
A. B. Balantekin,
``Character expansions, Itzykson-Zuber integrals, and
the QCD partition function,"
Phys. Rev. D {\bf 62} 085017/1-8 (2000).

\bibitem{Bal02}
A. B. Balantekin,
``Character expansions for the orthogonal and symplectic
groups,"
J. Math. Phys. {\bf 43} 604-620 (2002).

\bibitem{BS96}
A. N. Borodin and P. Salminen,
{\it Handbook of Brownian Motion
-- Facts and Formulae}, 2nd ed.
(Birkh\"auser, Basel, 2002).

\bibitem{Bru89}
M. F. Bru, 
``Diffusions of perturbed principal component analysis,"
J. Maltivated Anal. {\bf 29}, 127-136 (1989).

\bibitem{Bru91}
M. F. Bru, 
``Wishart processes,"
J. Theoret. Probab. {\bf 4}, 725-751 (1991).

\bibitem{deG68}
P.-G. de Gennes,
``Soluble model for fibrous structures with steric constraints,"
J. Chem. Phys. {\bf 48}, 2257-2259 (1968).

\bibitem{Doo84}
J. L. Doob,
{\it Classical Potential Theory and its Probabilistic Counterpart},
(Springer, New York, 1984).

\bibitem{Dys62a}
F. J. Dyson, 
``A Brownian-motion model for the eigenvalues of a random matrix,"
J. Math. Phys. {\bf 3}, 1191-1198 (1962).

\bibitem{Dys62b}
F. J. Dyson, 
``The threefold way. Algebraic structure of symmetry groups
and ensembles in quantum mechanics,"
J. Math. Phys. {\bf 3}, 1199-1215 (1962). 

\bibitem{Ede97}
A. Edelman,
``The probability that a random real Gaussian matrix has
$k$ real eigenvalues, related distributions, and
the circular law,"
J. Multivariate Anal. {\bf 60}, 203-232 (1997).

\bibitem{Efe97}
K. Efetov,
{\it Supersymmetry in Disorder and Chaos},
(Cambridge University Press, Cambridge, 1997).

\bibitem{EG95} 
J. W. Essam and A. J. Guttmann, 
``Vicious walkers and directed polymer networks in
general dimensions,"
Phys. Rev. E {\bf 52}, 5849-5862 (1995).

\bibitem{Fis84}
M. E. Fisher, 
``Walks, walls, wetting, and melting,"
J. Stat. Phys. {\bf 34}, 667-729 (1984).

\bibitem{FNH99}
P. J. Forrester, T. Nagao, and G. Honner,
``Correlations for the orthogonal-unitary and
symplectic-unitary transitions at the
hard and soft edges,"
Nucl. Phys. B {\bf 553[PM]}, 601-643 (1999).

\bibitem{Ful97}
W. Fulton,
{\it Young Tableaux with Applications to Representation
Theory and Geometry},
(Cambridge University Press, Cambridge, 1997).

\bibitem{FH91}
W. Fulton and J. Harris, 
{\it Representation Theory, A First Course},
(Springer, New York, 1991).

\bibitem{Gil74}
R. Gilmore,
{\it Lie Groups, Lie Algebras, and Some of Their Applications},
(John Wiley and Sons, New York, 1974).

\bibitem{Gin65}
J. Ginibre,
``Statistical ensembles of complex, quaternion, and real
matrices,"
J. Math. Phys. {\bf 6}, 440-449 (1965).

\bibitem{Gra99}
D. J. Grabiner,
``Brownian motion in a Weyl chamber, non-colliding particles,
and random matrices,"
Ann. Inst. Henri Poincar\'e,
Probab. Statist. {\bf 35}, 177-204 (1999).

\bibitem{HC57}
Harish-Chandra,
``Differential operators on a semisimple Lie algebra,"
Am. J. Math. {\bf 79}, 87-120 (1957).

\bibitem{Hel78}
S. Helgason,
{\it Differential Geometry, Lie Groups, and Symmetric Spaces},
(Academic, New York, 1978).

\bibitem{Hua63}
L. Hua,
{\it On the theory of functions of several complex variables. I},
tr. L. Ebner and A. Kor\'ani, 
(American Mathematical Society, Province, RI, 1963).

\bibitem{Imh84}
J. P. Imhof,
``Density factorizations for Brownian motion, 
meander and the three-dimensional
Bessel processes, and applications,"
J. Appl. Prob. {\bf 21}, 500-510 (1984).

\bibitem{IZ80}
C. Itzykson and J.-B. Zuber,
``The planar approximation. II,"
J. Math. Phys. {\bf 21}, 411-421 (1980).

\bibitem{JSV96}
A. D. Jackson, M. K. Sener and J. J. M. Verbaarschot,
``Finite volume partition functions and
Itzykson-Zuber integrals,"
Phys. Lett. {\bf B387}, 355-360 (1996).

\bibitem{KM59a}
S. Karlin and J. McGregor,
``Coincidence properties of birth and death processes,"
Pacific J. {\bf 9}, 1109-1140 (1959).

\bibitem{KM59b}
S. Karlin and J. McGregor,
``Coincidence probabilities,"
Pacific J. {\bf 9}, 1141-1164 (1959).

\bibitem{KNT03}
M. Katori, T. Nagao, and H. Tanemura,
``Infinite systems of non-colliding Brownian particles,"
Adv. Stud. in Pure Math. {\bf 39}
``{\it Stochastic Analysis on Large Scale Interacting Systems}",
283-306 (2004), (Mathematical Society of Japan, Tokyo);
arXiv:math.PR/0301143.

\bibitem{KT02}
M. Katori and H. Tanemura,
``Scaling limit of vicious walks and two-matrix model,"
Phys. Rev. E {\bf 66}, 011105/1-12 (2002).

\bibitem{KT03a}
M. Katori and H. Tanemura,
``Functional central limit theorems for vicious walkers,"
Stoch. Stoch. Rep. {\bf 75}, 369-390 (2003); 
arXiv:math.PR/0203286.

\bibitem{KT03b}
M. Katori and H. Tanemura,
``Noncolliding Brownian motions and Harish-Chandra
formula,"
Elect. Comm. in Probab. {\bf 8}, 112-121 (2003).

\bibitem{KT04}
M. Katori and H. Tanemura, in preparation.

\bibitem{KTNK03}
M. Katori, H. Tanemura, T. Nagao and N. Komatsuda,
``Vicious walk with a wall, noncolliding meanders, and
chiral and Bogoliubov-de Gennes random matrices,"
Phys. Rev. E  {\bf 68}, 021112/1-16 (2003).

\bibitem{KO01}
W. K\"onig and N. O'Connell,
``Eigenvalues of the Laguerre process as non-colliding 
squared Bessel process,"
Elec. Comm. in Prob. {\bf 6}, 107-114 (2001).

\bibitem{KGV00}
C. Krattenthaler, A. J. Guttmann, and X. G. Viennot, 
``Vicious walkers, friendly walkers and Young tableaux: II.
With a wall,"
J. Phys. A: Math. Gen. {\bf 33}, 8835-8866 (2000).

\bibitem{Mac82}
I. G. Macdonald,
``Some conjectures for root systems,"
SIAM J. Math. Anal. {\bf 13}, 988-1007 (1982).

\bibitem{Mac95}
I. G. Macdonald,
{\it Symmetric Functions and Hall Polynomials}, 2nd ed.
(Oxford Univ. Press, Oxford, 1995)

\bibitem{Meh91}
M. L. Mehta, 
{\it Random Matrices}, 2nd ed.
(Academic Press, London, 1991).

\bibitem{MP83}
M. L. Mehta and A. Pandey,
``On some Gaussian ensemble of Hermitian matrices,"
J. Phys. A: Math. Gen. {\bf 16}, 
(1983), 2655-2684.

\bibitem{Nag01}
T. Nagao, 
``Correlation functions for multi-matrix models and
quaternion determinants,"
Nucl. Phys. B {\bf 602}, 622-637 (2001).

\bibitem{Nag03}
T. Nagao,
``Dynamical correlations for vicious random walk with
a wall,"
Nucl. Phys. {\bf B658[FS]}, 373-396 (2003).

\bibitem{NF99}
T. Nagao and P. J. Forrester,
``Quaternion determinant expressions for multilevel
dynamical correlation functions of
parametric random matrices,"
Nucl. Phys. B {\bf 563[PM]}, 547-572 (1999).

\bibitem{NKT03}
T. Nagao, M. Katori, and H. Tanemura,
``Dynamical correlations among vicious random walkers,"
Phys. Lett. {\bf A307}, 29-35 (2003).

\bibitem{NRW86}
J.R. Norris, L.C.G. Rogers and D. Williams,
``Brownian motions of ellipsoids,"
Trans. Amer. Math. Soc. {\bf 294}, 757-765 (1986).

\bibitem{PM83}
A. Pandey and M. L. Mehta,
``Gaussian ensembles of random Hermitian matrices intermediate between
orthogonal and unitary ones,"
Commun. Math. Phys. {\bf 87}, 449-468 (1983).

\bibitem{RY98}
D. Revuz and M. Yor, 
{\it Continuous Martingales and Brownian Motion}, 3rd ed.
(Springer, New York, 1998).

\bibitem{SI03}
T. Sasamoto and T. Imamura,
``Fluctuations of the one-dimensional polynuclear growth model 
in a half space,"
J. Stat. Phys. {\bf 115}, 749-803 (2004).

\bibitem{Sel44}
A. Selberg,
``Bemerkninger om et multiplet integral,"
Norsk Matematisk Tidsskrift
{\bf 26}, 71-78 (1944).

\bibitem{SV98}
M. K. Sener and J. J. M. Verbaarschot,
``Universality in chiral random matrix theory at
$\beta=1$ and $\beta=4$,"
Phys. Rev. Lett. {\bf 81}, 248-251 (1998).

\bibitem{Sta99}
R. P. Stanley,
{\it Enumerative Combinatorics},
vol.2,
(Cambridge University Press, Cambridge, 1999).

\bibitem{Ver94}
J. Verbaarschot,
``The spectrum of the Dirac operator near zero
virtuality for $N_{c}=2$ and chiral random matrix theory,"
Nucl. Phys. B {\bf 426[FS]}, 559-574 (1994).

\bibitem{VZ93}
J. J. M. Verbaarschot and I. Zahed,
``Spectral density of the QCD Dirac operator near
zero virtuality,"
Phys. Rev. Lett. {\bf 70}, 3852-3855 (1993).

\bibitem{Yor92}
M. Yor,
{\it Some Aspects of Brownian Motion, 
Part I: Some Special Functionals},
(Birkh\"auser, Basel 1992).

\bibitem{Zir96}
M. R. Zirnbauer,
``Riemannian symmetric superspaces and their origin
in random-matrix theory,"
J. Math. Phys. {\bf 37}, 4986-5018 (1996).

\end{thebibliography}
\end{document}